\documentclass[a4paper,12pt,english]{article}
\pdfoutput=1

\usepackage[bindingoffset=0.3cm,textheight=22.5cm,hdivide={2.7cm,*,2.7cm}, vdivide={*,22cm,*}]{geometry}
\usepackage[bookmarksnumbered=true,breaklinks=true]{hyperref}

\usepackage{amsmath,amsfonts,amssymb,babel,slashed,graphicx,color,empheq}

\usepackage[numbers,square,comma,sort&compress]{natbib}

\usepackage[toc,page]{appendix}
\newcommand{\stoptocwriting}{%
  \addtocontents{toc}{\protect\setcounter{tocdepth}{-5}}}
\newcommand{\resumetocwriting}{%
  \addtocontents{toc}{\protect\setcounter{tocdepth}{\arabic{tocdepth}}}}

\makeatletter

\newcommand{\e}{\mathrm{e}}

\newcommand{\bbm}{\left(\begin{matrix}}
\newcommand{\ebm}{\end{matrix}\right)}
\newcommand{\beq}{\begin{eqnarray}}
\newcommand{\eeq}{\end{eqnarray}}
\makeatother

\newcommand{\del}{\partial}

\newcommand{\be}{\begin{equation}}
\newcommand{\ee}{\end{equation}}

\newcommand{\beqa}{\begin{eqnarray}}
\newcommand{\eeqa}{\end{eqnarray}} \newcommand{\eq}[1]{(\ref{#1})}
\def\nn{\nonumber} \def \bea{\begin{eqnarray}} \def\eea{\end{eqnarray}}

\newcommand{\barr}{\begin{array}}
\newcommand{\earr}{\end{array}}
\numberwithin{equation}{section}

\def\a{\alpha}  \def\b{\beta}
 \def\g{\gamma} 
 \def\d{\delta} \def\D{\Delta}
    \def\k{\kappa}
\def\l{\lambda} \def\L{\Lambda}  
    \def\r{\rho}
\def\s{\sigma}  \def\t{\tau}

\def\cA{{\cal A}}  \def\cC{{\cal C}} 
  \def\cF{{\cal F}} 
\def\cG{{\cal G}} \def\cH{{\cal H}}  
 \def\cK{{\cal K}}  
\def\cM{{\cal M}}  \def\cO{{\cal O}} 
\def\cP{{\cal P}}  \def\cR{{\cal R}} 
\def\cS{{\cal S}} \def\cT{{\cal T}}

\def\R{{\mathbb R}} \def\C{{\mathbb C}} 

 \def\one{\mbox{1 \kern-.59em {\rm l}}}

\def\mmu{\mathfrak{u}}

\def\msu{\mathfrak{su}}
\def\mso{\mathfrak{so}}

\def\A{{\bf A}}

\def\bit{\begin{itemize}} \def\eit{\end{itemize}} \def\Tr{\mbox{Tr}}

\def\({\left(} \def\){\right)}

 \allowdisplaybreaks[3]

\begin{document}

\makeatother


\parindent=0cm

\renewcommand{\title}[1]{\vspace{10mm}\noindent{\Large{\bf

#1}}\vspace{8mm}} \newcommand{\authors}[1]{\noindent{\large

#1}\vspace{5mm}} \newcommand{\address}[1]{{\itshape #1\vspace{2mm}}}


\begin{titlepage}
\begin{flushright}
 UWThPh-2016-8 
\end{flushright}
\begin{center}
\title{ {\Large Emergent gravity on  covariant quantum spaces \\[1ex]
 in the IKKT model }  }

\vskip 3mm

\authors{Harold C. Steinacker{\footnote{harold.steinacker@univie.ac.at}}
}
 
\vskip 3mm

 \address{ 

{\it Faculty of Physics, University of Vienna\\
Boltzmanngasse 5, A-1090 Vienna, Austria  }  
  }

\bigskip

\vskip 1.4cm

\textbf{Abstract}
\vskip 3mm

\begin{minipage}{14cm}%

We study perturbations of 4-dimensional fuzzy spheres as  backgrounds in the IKKT or IIB  matrix model. 
Gauge fields and metric fluctuations are identified among the excitation modes with lowest spin, 
supplemented by a  tower of higher-spin fields.
They arise from an internal structure which can be viewed as a twisted bundle over $S^4$,
leading to a covariant noncommutative geometry. 
The linearized 4-dimensional Einstein equations are obtained 
from the classical matrix model action under certain conditions,  modified by an IR cutoff.
Some one-loop contributions to the effective action are computed using the formalism of string states.

\end{minipage}

\end{center}

\end{titlepage}

\tableofcontents

\section{Introduction}

Matrix models such as the IIB or IKKT model \cite{Ishibashi:1996xs} 
(cf. \cite{Banks:1996vh,deWit:1988wri}) 
provide fascinating candidates for
a quantum theory of fundamental interactions. Their most interesting feature is that
geometry is not an input, but arises itself as a brane-type solution with dynamical ``quantum'' geometry.
Fluctuations around such solutions  lead to gauge fields and matter fields on the background. 
It is  natural to expect that 
gravity, along with the other fundamental interactions, should  emerge on suitable backgrounds
in the low-energy, semi-classical regime.
Remarkably, numerical evidence 
for the emergence of 3+1-dimensional space-time within the finite-dimensional IIB model was reported 
recently \cite{Kim:2011cr,Kim:2012mw}.

At first sight,
the relation of the IIB matrix model with string theory 
suggests that 4-dimensional gravity can arise only if target space is compactified. This would not only 
lead to the well-known issues with a vast landscape of possibilities, it would also require ad-hoc 
modifications or constraints\footnote{For example, the toroidal compactifications considered in 
\cite{Connes:1997cr} require an infinite number of degrees of freedom and cannot be imposed in the 
finite-dimensional matrix model.} of the matrix model, destroying much of its appeal and simplicity.
With this motivation, there were ongoing efforts to understand  possible mechanisms for gravity in this model 
 based solely on the 4-dimensional, non-commutative (NC) physics of the branes rather than
the 10-dimensional bulk gravity (which arises in the matrix model upon quantization) 
\cite{Steinacker:2007dq,Steinacker:2010rh,Yang:2006dk,Yang:2009pm}. 
Although 4-dimensional NC gauge theory behaves indeed very much like a gravitational theory 
\cite{Rivelles:2002ez,Yang:2006dk,Szabo:2006wx,Steinacker:2010rh,Szabo:2009tn}, 
the emerging gravity on basic branes  seems to be  different from usual gravity,
and it was not possible to derive the Einstein equations up to now.

In this paper, we show that  Einstein-like gravity can indeed arise on more sophisticated, {\em covariant} 
noncommutative branes in this model, at least in some regime.  
This is based solely on the classical matrix model dynamics for fluctuation modes on the background brane, and 
has {\em nothing} to do with IIB supergravity in the bulk. 
The internal structure of the quantum space  is crucial for the mechanism. This background is a generalized
4-dimensional fuzzy sphere $\cS^4_\L$, but most of the considerations should apply also to analogous spaces
with Minkowski signature. 

There are two crucial features of $\cS^4_\L$ which are essential here \cite{Ho:2001as}. First, it 
has an internal bundle structure, which transforms  non-trivially under local space(time) rotations. 
Each point on the local fiber corresponds to a particular choice of 
an antisymmetric tensor $\theta^{\mu\nu}$ on $S^4$.
This tensor is averaged  over the fiber, leading to a covariant noncommutative structure of the 
4-dimensional space. 
The second crucial feature is the fact that  $\theta^{\mu\nu}$ (complemented by $P^\mu$)
is not central, but generate the local Euclidean isometry group including translations. 
Quantum spaces with these features will be denoted as {\em covariant}
quantum spaces. This concept is actually very old and goes back to
Snyder  and his proposal \cite{Snyder:1946qz} for a Lorentz-invariant noncommutative Minkowski space.
Fuzzy $S^4$ is a compact and well-controlled Euclidean version of such a space.
Due to the extra generators $\theta^{\mu\nu}$ and $\cP^\mu$, 
the corresponding algebra (of ``functions``) 
is larger than what seems to be needed in field theory, hence  
this type of space was not very much appreciated.

In contrast to most previous work on this type of  spaces 
(cf. \cite{Battisti:2010sr,Girelli:2010wi} and references therein),
we take serious these extra fluctuation modes.
They can be understood as harmonics on the internal fiber, which 
-- in contrast to  Kaluza-Klein compactification --
transform non-trivially under the local isometry group. This  leads to 
an  infinite tower of higher-spin fields, truncated at $N$ for fuzzy $\cS^4_N$. 
Among the lowest modes in this tower, we  identify 
the metric fluctuation, a selfdual $SO(4)$ connection as well as gauge fields. 
We then perform a fluctuation analysis  in the semi-classical limit along the lines of 
\cite{Steinacker:2007dq,Steinacker:2010rh}. 
%
The metric fluctuation $H_{\mu\nu}$ is a combination of a rank 2 tensor 
field $h_{\mu\nu}$ and the divergence $\del^\r A_{\mu\r\nu}$ of a $SO(4)$-valued gauge field.
Their semi-classical equations of motion of the classical matrix model then 
lead to the (linearized) Einstein equations for $H_{\mu\nu}$, for length scales below some IR cutoff scale $m^{-1}$.
Above this length scale, gravity no longer applies. 
However, this  requires a certain type of generalized fuzzy spheres $\cS^4_\L$, and
we  have to assume dimensional reduction to 4 dimensions.
Mechanisms to ensure this are suggested,
 but this needs to be addressed in future work.
This issue could be avoided by 
a suitable self-dual modification of the matrix model action.

The present framework incorporates several aspects of previous work in this context. 
Averaging over the Poisson structure $\theta^{\mu\nu}$ was considered in 
the DFR approach to field theory on the Moyal-Weyl plane  \cite{Doplicher:1994tu}, in order to preserve Lorentz invariance.
However, $\theta^{\mu\nu}$ was considered as central there, which kills gravity.
The effective metric and the dynamics of NC branes in matrix models was analyzed in \cite{Steinacker:2007dq,Steinacker:2010rh}, 
but the backgrounds under consideration were too simple. Finally, an interpretation of the matrices as covariant 
derivatives rather than position operators was proposed in \cite{Hanada:2005vr}. 
This also leads to higher spin fields with some similarities to the present framework and even
the Einstein equations in vacuum, however this doesn't work in the finite-dimensional model, and the proper 
coupling to matter was not established. 
Due to the $SO(5)$ setup, there
are also similarities with the MacDowell-Mansouri formulation of GR \cite{MacDowell:1977jt,Wise:2006sm},
however the physics is different: the full $SO(5)$ symmetry is manifest here, and there are additional 
degrees of freedom beyond the ones in GR.
The present framework shares aspects with noncommutative $SO(5)$ gauge theory approaches
\cite{Chaichian:2007we,Chamseddine:2000si,Cardella:2002pb}, 
but again this is not quite appropriate: the gauge group is actually much larger here, corresponding 
to (a quotient of) $U(\mso(5))$.

The  reason for insisting on the IIB model is that the quantization 
is well-behaved, since the non-local UV/IR mixing is mild due to maximal SUSY. 
In section \ref{sec:one-loop}, we compute 
 the leading terms in the one-loop effective action for the 
lowest fluctuation modes on $\cS^4_N$. 
This is possible due to recent progress for the quantization of field theory 
on fuzzy spaces based on string states \cite{Steinacker:2016xx}. 
We show that previous one-loop results  can be reproduced efficiently in this formalism, and 
some (preliminary) computations suggest that the one-loop effects can be captured by a minor generalization
of the classical action, preserving the  mechanism for gravity.

This paper is  written in a pedestrian way,  to make everything explicit 
and to avoid getting trapped in some formalism. Of course there should be a more structural approach, 
and many limitations of this paper - notably the restriction to the linearized regime - are clearly inessential.
Other open issues include the coupling of
the conformal mode to scalar fields which seems odd (see section \ref{sec:coupl-scalar-matter}), 
the proper extension to the Minkowski case, the justification of dimensional reduction for 
the generalized sphere, and
the coupling to fermions. These should be addressed in future work.
Nevertheless the basic mechanism is  compelling, 
and should provide a serious candidate for a quantum theory of gravity
which behaves similar to GR in a suitable range.

\section{Covariant fuzzy four-spheres $\cS^4_\L$}

We consider covariant fuzzy four-spheres defined 
in terms of 5  hermitian matrices $X^a, \ a=1,..., 5$
acting on some finite-dimensional Hilbert space $\cH$, 
which transform as vectors under $SO(5)$ 
\begin{align}
 [\cM_{ab},X_c] &= i(\d_{ac} X_b - \d_{bc} X_a), \nn\\
  [\cM_{ab},\cM_{cd}] &=i(\d_{ac}\cM_{bd} - \d_{ad}\cM_{bc} - \d_{bc}\cM_{ad} + \d_{bd}\cM_{ac}) \ .
 \label{M-M-relations}
\end{align}
Here the $\cM^{ab}=- \cM^{ba}$ for $1\leq a\neq b \leq 5$ define a (not necessarily irreducible) representation of $\mso(5)$ on $\cH$. 
The radius 
\begin{align}
 X^a X_a &= \cR^2 
\end{align}
is a scalar operator of dimension $L^2$, and the commutator of the $X^a$ will be denoted by 
\begin{align}
  [X^a,X^b]  &=: i \Theta^{ab} \ .
\end{align}
Here and throughout this paper, indices are raised and lowered with  $g_{ab} = \d_{ab}$.
This type of relations constitute a {\em covariant quantum space}.

The form of the algebra \eq{M-M-relations} suggests a particular realization of
such fuzzy four-spheres, based on an irreducible representation (irrep) of 
$\mso(6)$ as follows\footnote{This is similar to an observation of Yang \cite{Yang:1947ud} in the context of the 
Snyder's noncommutative space.}
\begin{align}
 X^a &= r \cM^{a6}, \qquad a = 1,...,5 \ , \qquad \Theta^{ab} = r^2 \cM^{ab}
\end{align}
Here $\cM^{ab}, \ a = 1,...,6$ define an irrep of $\mso(6) \cong \msu(4)$ on $\cH$,
and $r$ is a scale parameter of dimension $L$. 
Correspondingly, $\mso(5)\subset \mso(6)$ is embedded
by restricting the indices of $\cM^{ab}$ to $a,b=1,...,5$.
We also note the following simple identity for such spheres 
\begin{align}
 \{X_a,\Theta^{ab}\}_+ &= [\cR^2,X^b]  \ \neq 0 \qquad \mbox{in general} . 
 \label{MM-exact-1}
\end{align}
This is the type of space under consideration in this paper. There are important differences 
depending on the representation $\cH$ of $\mso(6)$:

\paragraph{The basic fuzzy 4-sphere $\cS^4_N$.}

The simplest example is the ''basic`` fuzzy four-sphere 
$\cS^4_N$ \cite{Grosse:1996mz,Castelino:1997rv,Ramgoolam:2001zx},
which is obtained for the highest weight irrep  $\cH = \cH_\L$ of $\mso(6)$ with $\L = (0,0,N)$, 
denoting highest weights by their Dynkin indices. 
This representation  can be realized as  
totally symmetric tensor product $\cH_\L \cong (\C^4)^{\otimes_S N}$
of the 4-dimensional (spinor) representation of $\mso(6)$, which happens to remain 
irreducible as a $\mso(5) \subset \mso(6)$ representation.
In this particular case, the radius operator is proportional to the identity operator,
 \begin{align}
  X^a X_a &= \cR^2 = r^2 R_N^2 \one, \qquad  R_N^2 = \frac 14 N(N+4) \ 
\end{align}
For this basic fuzzy sphere $\cS^4_N$, the following useful formulae 
are established in appendix \ref{sec:S4-properties}
\begin{align}
\{X_a,\cM^{ab}\}_+ &=  0  \nn\\
 \frac 12\{\Theta^{ab},\Theta^{a'c}\}_+ g_{aa'} &= r^2 \cR^2 \big(g^{bc} -\frac 1{2\cR^2} \{X^b, X^c\}_+\big) \label{ThTh-N}\\
 \epsilon^{ijklm} X_iX_jX_kX_l X_m &= (N+2) r^3 \cR^2
 \label{MM-exact}
\end{align}
where $\{.,.\}_+$ denotes the anti-commutator.
As explained in appendix \ref{sec:app-geom-general}, this is the 
quantization of a 6-dimensional coadjoint orbits of $SO(6)$ mapped to  $S^4\hookrightarrow \R^5$
via the $x^a\sim X^a$.

\paragraph{Generalized fuzzy 4-spheres $\cS^4_\L$.}

More general fuzzy 4-spheres are obtained for $\mso(6)$ irreps
$\cH_\L$ with $\L = (n_1,n_2,N)$.
As explained in appendix \ref{sec:app-geom-general}, 
these arise as quantizations of generic coadjoint $SO(6)$ orbits, and
the semi-classical geometry is that of a ''thick`` 4-sphere  embedded in $\R^5$.
As long as  $n_1, n_2 \ll N$,
the radius $\cR^2= X_a X^a$  is non-trivial but with sharply peaked spectrum around $r^2 R_N^2$,
with 
\begin{align}
 [\cR^2,X^b]  \ \neq 0 .
 \label{Rsquare-nontriv}
\end{align}
As explained in Appendix \ref{sec:app-geom-general},
 $\cS^4_{\L}$  can be understood as a $SU(3)$ bundle over $\cS^4_N$.
The  relation \eq{ThTh-N}  is modified as
\begin{align}
  \frac 12\{\Theta^{ab},\Theta^{a'c}\}_+ g_{aa'} &= r^2 \cR^2 \big(g^{bc} -\frac 1{2\cR^2} \{X^b, X^c\}_+ + t^{bc}\big) 
  \label{ThTh-generic}
\end{align}
where $t^{ab} = O\big(\frac{n}{N}\big)$, see \eq{MM-6-generic} and \eq{mm-id-generalS4}.
This  generalization will be essential for gravity.

\subsection{Semi-classical geometry and mode decomposition}

As usual for fuzzy or noncommutative spaces,
the matrix algebra $End(\cH)$ constitutes the noncommutative algebra of 
functions or fields on the (generalized) fuzzy 4-sphere. 
The realization of $\cM^{ab}$  in terms of generators  of $\mso(6)\cong \msu(4)$ 
also provides the proper geometrical interpretation.
We recall the well-known fact  that  $End(\cH_\L)$ can be naturally 
interpreted as quantized algebra of functions on the
coadjoint orbit $\cO[\L] = \{g\cdot \L \cdot g^{-1}; \ g\in SU(4)\}$
of $\msu(4)$ through the weight $\L$ (cf. \cite{Hawkins:1997gj}). The generators $\cM^{ab}, \ a,b=1,...,6$ are  
quantized embedding functions 
\begin{align}
 \cM^{ab} \sim m^{ab}: \quad \cO[\L] \hookrightarrow \R^{15} \cong \msu(4) \ 
\end{align}
dual to some ON basis $\l^{ab}$ of $\msu(4)$.
In particular, the
$X^a \sim \cM^{a6}$ are naturally interpreted as {\em projections} of such coadjoint orbits to $S^4  \ \subset\R^5$,
\begin{align}   \fbox{$
 X^a \sim x^a: \quad \cO[\L] \ \hookrightarrow \ \R^{15} \ \stackrel{\Pi}{\to} \  S^4  \ \subset \R^5 \ 
 $}
 \label{X-embed-Pi}
\end{align}
where  $\Pi$ denotes the projection of $\msu(4)$ to the subspace spanned by the $\l^{a6}$
generators. 
Hence the fuzzy 4-spheres are actually higher-dimensional homogeneous spaces which are twisted bundles over $S^4$, 
with the fiber playing the role of a
hidden extra dimension. In contrast to standard Kaluza-Klein compactifications, these extra dimensions 
lead to higher-spin modes here. 
For the basic  4-sphere $\cS^4_N$, the underlying orbit is $\cO[\L] = \C P^3$, which is a $S^2$ bundle over $S^4$
as elaborated in  \cite{Ho:2001as,Medina:2002pc,Fabinger:2002bk,Abe:2004sa,Karabali:2006eg,Medina:2012cs,Steinacker:2015dra}.
More details on the geometry including the generic case are given in Appendix \ref{sec:app-geom-general}.
In particular,
the space of classical functions on these orbits is spanned by polynomials 
$F(x^a,m^{ab})$, which are in one-to-one correspondence with the noncommutative modes \eq{NC-modes},
up to some UV cutoff defined by $N$ and $n_i$.

\paragraph{Poisson structure.}

This geometrical picture also explains the origin of the commutator as quantized (Kirillov-Kostant)
Poisson bracket on $\cO[\L]$.
This Poisson structure can be viewed as a 2-vector field on $\cO[\L]$
\begin{align}
 \{f,g\} = \theta^{AB}\del_A f \del_B g
\end{align}
whose projection (push-forward) to $S^4$ is given by 
\begin{align}
 \theta^{\mu\nu}(x,\xi)\del_\mu \otimes \del_\nu \ .
\end{align}
Here $\xi$ are coordinates on the internal fiber of $\cO[\L]$ over $S^4$.
For the basic 4-sphere $\cS^4_N$, $\theta^{\mu\nu}(x,\xi)$ is 
selfdual (SD) at each $(x,\xi)$, defining a bundle of  SD frames over $S^4$,
which rotates (and averages out) along the fiber $S^2$.
More precisely, it transforms as $(1,0)$ under the local $SO(4) = SU(2)_L\otimes SU(2)_R$ rotations, 
which are implemented by the  
action of
$\{\theta^{\mu\nu},.\}$ on itself. In the noncommutative case, this amounts to a gauge transformation
\begin{align}
\L^{\mu\mu'}  \L^{\nu\nu'} \Theta^{\mu'\nu'}
 = U_\L\Theta^{\mu\nu}U_\L^{-1} \ .
\end{align}
In other words, local rotations are implemented as gauge transformations, which already  
hints towards  gravity. 

For the 5 embedding functions $x^a \sim X^a$, the Poisson bracket
\begin{align}
 \{x^a,x^b\} =  \theta^{ab} :\quad \cO[\L] \hookrightarrow \mso(5) \subset \mso(6)
\end{align}
 gives rise to $\Theta^{ab}$ is the fuzzy case. Once again,
$\theta^{ab}$ is only defined on the bundle $\cO[\L]$, it 
is {\em not} a Poisson bracket on $S^4$, 
since it is not constant along the fiber\footnote{Recall that $H^2(S^4) = 0$, hence there 
is no symplectic 2-form on $S^4$.}.

Much of the analysis in this paper is done in this semi-classical limit indicated by  $\sim$, 
replacing commutators by Poisson brackets 
and working with $\cO[\L]$. This greatly simplifies the analysis, 
and it is certainly justified in the gravity regime where the typical wavelengths are much longer than the scale of noncommutativity.

\paragraph{Coherent states.}

As for all quantized coadjoint orbits, coherent states on $\cO[\L]$
are given by highest weight states $|\L\rangle\in \cH_\L$ and their $SO(6)$  orbits,
\begin{align}
 |{\bf x}\rangle &\equiv |x;\xi\rangle = g_{\bf x} \cdot |\L\rangle, \qquad g_{\bf x} \in SO(6)  \nn\\[1ex]
 x^a &=  \langle{\bf x}| X^a |{\bf x}\rangle \equiv \langle X^a\rangle \ .
\end{align}
Up to a $U(1)$ phase factor, they are in one-to-one correspondence to 
points ${\bf x}$ on $\cO[\L]$. Alternatively, we can use the $SO(5)$ point of view and consider
highest weight states of the $SO(5)$ modules. It will suffice here to consider the case of the basic 
fuzzy sphere $\cS^4_N$, where both notions coincide. We can then
 label the points on the bundle $\cO[\L] \cong \C P^3$ locally by $x\in S^4$ and $\xi$, where 
the ``north pole`` ${\bf p}$ corresponds to the highest weight state $|\L\rangle$.
They are optimally localized, minimizing the uncertainty 
in position space\footnote{Here  $\langle{\bf p} |X^5| {\bf p}\rangle = \frac {r}2 N$ at the north pole $p$ follows
using the explicit realization of $X^a$ in terms of gamma matrices \cite{Castelino:1997rv}. }
\begin{align}
 \Delta^2 \ &:= \sum_a\langle(X^{a} - \langle X^a\rangle)^2\rangle 
  = \sum_a\langle (X^a)^2\rangle  - \langle  X^a \rangle^2 \nn\\
  &= (R_N^2 - \frac 14 N^2) r^2 
  \sim \frac{4}{N} \cR^2  = 2 r \cR \nn\\
  &=: L_{NC}^2 \ 
 \label{Delta-S4}
\end{align}
which defines the length scale $L_{NC}$.
One can then associate to any operator  $\phi\in End(\cH)$ a function
$\phi({\bf x})$ on $\cO[\L]$ as follows 
\begin{align}
 \phi({\bf x}) = \langle {\bf x}| \phi |{\bf x}\rangle, 
\end{align}
and the semi-classical regime is characterized by functions $\phi(x)$ which vary on scales  $> L_{NC}$.
The coherent states form a $U(1)$ bundle over $\cO[\L]$, with a canonical
connection whose curvature gives the symplectic form $\omega$ on $\cO[\L]$, corresponding to the Poisson structure 
\begin{align}
  i\theta^{ab}({\bf x}) =  \langle {\bf x}| [X^a,X^b]|{\bf x}\rangle \ .
\end{align}
This also encodes the uncertainty scale $L_{NC}$ and  the volume quantization via \eq{MM-exact}.
Finally, the trace over $End(\cH)$ can be realized by the integral over all coherent states on $\cO[\L]$, 
\begin{align}
 Tr \phi = \frac{\dim\cH_\L}{Vol \cO[\L]}\int_{\cO} d{\bf x} \langle {\bf x}| \phi |{\bf x}\rangle \ .
 \label{trace-coherent}
\end{align}
This locally separates into an integration over $S^4$ times the internal fiber $\cF$,
which allows to evaluate the matrix model actions in a standard semi-classical form.

\paragraph{Scalar fields and higher-spin modes.}

The most general functions on fuzzy $\cS^4_N$ are organized into the 
following   $SO(6)$ resp. $SO(5)$ modes (cf. \cite{Ramgoolam:2001zx,Medina:2002pc,Medina:2012cs,Steinacker:2015dra})
\begin{align}
 \phi \in  End(\cH) \cong  \bigoplus\limits_{n\leq N} (n,0,n)_{\mso(6)} 
 \cong \bigoplus\limits_{m\leq n\leq N} (n-m,2m)_{\mso(5)}  \ ,
 \label{mode-expansion-alg}
\end{align}
denoting highest weight irreducible representations (irreps) by their Dynkin indices;
for example, $(n,m)$ denotes the $\mso(5)$ irrep 
with highest weight $\L = n\L_1 + m\L_2$ where $\a_1$ is the long root and $\a_2$ the short root.
We are mainly interested in  the ``low spin'' representations with small $m$.
Then a more explicit realization is obtained 
in terms of ordered polynomials\footnote{This is a quotient of the Poincare-Birkhoff-Witt basis of 
$U(\mso(6))$.} 
in the generators $X^a$ and $\cM^{ab}, \ a,b=1,...,5$. 
For example, scalar fields on $S^4$ correspond to the $(n,0)$ modes, realized by
totally symmetric polynomials $F(X) = F_{a_1 ... a_n} X^{a_1} ... X^{a_n}$,
and denoted by
\begin{align}
 \cC_N(S^4) := \bigoplus \limits_{n\leq N} F_n(X) \ \cong  \bigoplus\limits_{n\leq N} (n,0) \ . 
 \label{scalar-fields}
\end{align}
Then
\begin{align}
  \phi({\bf x}) = \langle {\bf x}| \phi |{\bf x}\rangle , \qquad \phi \in \cC_N(S^4)
\end{align}
is constant along the fiber and defines a function on $S^4$. 
There is an associated projection map \cite{Ramgoolam:2001zx}  
\begin{align}
 \phi \mapsto  [\phi]_{0} := \phi_0  \ \in \cC_N(S^4) ,
 \label{average-def}
\end{align}
which picks out the scalar  modes $(n,0)$  in \eq{mode-expansion-alg}.
In the semi-classical limit, this corresponds to integrating $\phi({\bf x})$ over the internal fiber.

 More generally, we can
organize all other higher spin fields in terms of polynomials with 
''internal`` generators $\cM^{ab}, \ a,b=1,...,5$ multiplied by scalar functions.
For example, 
\begin{align}
 F_{ab}(X) \cM^{bc} &= F_{a_1 ... a_n;bc} X^{a_1} ... X^{a_n}\cM^{bc} \quad \in \ \ (n+1,2)_{\mso(5)} \nn\\
 F_{bc;de}(X) \cM^{bc} \cM^{de} &= F_{a_1 ... a_n;bc;de} X^{a_1} ... X^{a_n}\cM^{bc} \cM^{de}
 \ \in \ \ (n+2,4)_{\mso(5)}
 \label{NC-modes}
\end{align}
and so forth, where
the $F_{a_1 ... a_n;bc}$ and $F_{a_1 ... a_n;bc;de}$ are  tensors of $SO(5)$
corresponding to Young tableaux with one row of length 2 and two rows of length 2, respectively.
In particular, the $ F_{bc}(X) \cM^{bc}$ can be identified with 
2-forms $F_{bc}(x) dx^b \wedge dx^c$ on $\R^5$.
These $(n,m)$ modes with $m\neq 0$ correspond to functions on $\cO[\L] \cong \C P^3$ which are non-trivial 
harmonics on the $S^2$ fiber. They are higher-spin fields on $S^4$
rather than Kaluza-Klein modes, because the 
local Lorentz group acts non-trivially on the internal $S^2$ fiber. 
This  leads to a  higher-spin theory, and we will show that its
spin 2 sector  describes  gravity, but only for the  generic spheres $\cS^4_\L$.

For the generalized spheres $\cS^4_\L$,
the scalar  operator $\cR^2 = X_a X^a$ is a non-trivial $\mso(5)$ Casimir operator which
distinguishes some of the internal structure.
Then the mode decomposition  
is analogous but more complicated, with  multiplicities arising in 
the decomposition \eq{mode-expansion-alg}. E.g. for $\L = (k,0,N)$, one finds schematically 
\begin{align}
  End(\cH_\L) \cong  \bigoplus\limits_{n\leq N} k_n(n,0,n) \ \oplus ... \ \mbox{(other modes)}\ . 
 \label{mode-expansion-alg-generic}
\end{align}

\subsection{Local description}

We would like to understand the local structure from 
a field theory point of view, near some reference point $p\in S^4$ denoted as ``north pole''.
We pick a coherent state $|{\bf p}\rangle$ to mark this point.
Throughout this paper, tensorial objects at the (arbitrary) point $p\in S^4$   will be expressed 
in terms of the local tangent space $T_p S^4$, using the 4 tangential Cartesian coordinates $x^\mu$ centered at $p$, 
with
\begin{align}
 x^\mu(p) = \langle{\bf p}| X^\mu |{\bf p}\rangle &= 0, 
 \qquad  \langle{\bf p}| X^5 |{\bf p}\rangle = \frac{rN}{2} =: R \approx \cR \ 
\end{align}
assuming $n_i \ll N$.
Then quantities such as $x^\mu(p)$ can always be dropped, greatly simplifying the analysis.
Thus we can view $x^\mu$ as Riemannian normal coordinates at $p$ with respect to the 
embedding metric $g_{\mu\nu}$ of $S^4 \subset \R^5$, and $\nabla^{[g]}_\mu|_p = \del_\mu|_p$. 
To avoid confusions with the effective gravitational metric, we will use the symbol $\del_\mu$ for $\nabla^{[g]}_\mu$,
and we will often drop its (``cosmologically small'') 
curvature $[\nabla^{[g]}_\mu,\nabla^{[g]}_\nu] = O(\frac 1{R^2})$ for simplicity.
The generators separate accordingly as 
\begin{align}
 X^a = \begin{pmatrix}
        X^\mu \\ X^5
       \end{pmatrix}  ,  
\end{align}
with $X^5 = \sqrt{R^2 - X_\mu X^\mu}$,
and the 4 matrices $X^\mu \sim x^\mu$ are quantizations of these local coordinates. 
The stabilizer of $p$ (or $X^5$) is given by $SO(4)$.
Accordingly,  $\mso(5)$  decomposes into $\mso(4)$ and local ``translation`` generators,
\begin{align}
 \cM^{ab} = \begin{pmatrix}
             \cM^{\mu\nu} &  \cP^\mu \\
             -\cP^\mu & 0
            \end{pmatrix} \qquad \mbox{where} \quad \cP^\mu = \cM^{\mu 5} \ .
\end{align}
In this setup,  the (Euclidean) Poincare-group  $ISO(4)$  is recovered as usual by a 
contraction 
\begin{align}
 P_\mu &= \frac 1{R} g_{\mu\nu} \cP^\nu,  \qquad  X^\mu = r Z^\mu, \qquad R = R_N r 
 \end{align}
  taking $R$ to be much larger than any other length scale under consideration.
 Then the $\cS^4_\L$ algebra takes the form
 \begin{align}
 [P_\mu,X^\nu] &= i \frac{X^5}{R}\,\d_\mu^{\nu} \quad \ \ \stackrel{R\to\infty}{\to} \ i \d_\mu^{\nu},  \nn\\ 
 [P_\mu,P_\nu] &= \frac i{R^2} \cM^{\mu\nu} \quad  \stackrel{R\to\infty}{\to} \ 0  \nn\\
 [X^\mu,X^\nu] &=: i\theta^{\mu\nu} =  i r^2  \cM^{\mu\nu} 
\end{align}
assuming  $r \ll 1$.
Here and in the following, greek indices  indicate that the corresponding tensor is  tangential.

\paragraph{Poisson algebra limit.}

Now consider again the semi-classical (Poisson) limit.
The exact relations  \eq{ThTh-N} for $\cS^4_N$ then imply the following important
formula 
\begin{align}
 g_{cc'}\theta^{c a} \theta^{c' b} \ &= \ \frac 14 \D^4\,P_T^{ab}, \qquad 
 P_T^{ab} = g^{ab} - \frac 1{R^2}\,x^a x^b, \nn\\
 P_T^2 &= P_T , \qquad P_T \cdot x = 0 \ .
 \label{MM-formula-nonloc}
\end{align}
where $P_T^{ab}$  is the projector on the tangent space of $S^4 \subset \R^5$.
This allows to evaluate the {\bf kinetic term of a scalar field} $\phi\in\cC_N(S^4)$ \eq{scalar-fields} 
in the semi-classical limit:
\begin{align}
 -g_{ab}[X^a,\phi][X^b,\phi] 
 &\sim  g_{ab}\theta^{a\mu'} \theta^{b\nu'}\del_{\mu'}\phi \del_{\nu'}\phi  
  =  \g^{\mu\nu} \del_{\mu}\phi \del_{\nu}\phi \ .
 \end{align}
 As always, this is obtained by replacing commutators with Poisson brackets. 
  Here
\begin{align}
 \g^{\mu\nu} &:=  g_{ab}\theta^{a\mu} \theta^{b\nu}  =  \frac 14 \D^4\, g^{\mu\nu}  \ 
 \label{MM-formula-metric}
\end{align}
will play a prominent role as effective background metric. 
 In contrast to $\theta^{\mu\nu}$, 
 this is indeed a tensor on $S^4$, i.e. it is constant along  the fiber.
 For the translation generators $p_\mu = \frac{1}{R r^2}\theta^{\mu 5}$,
equation \eq{MM-exact} implies
\begin{align}
  \theta^{\mu\nu} p_\nu = \frac{1}{Rr^2} \theta^{\mu\nu} \theta_{\nu 5} 
  = \frac {x^5}{R} \, x^\mu \ . 
 \label{theta-P-exact}
\end{align}
This   shows that  on $\cS^4_N$, the functions $p_\mu$ are not independent
and basically vanish, since $x_\mu=0$ in the local frame at any given point $p\in S^4$.
See appendix \ref{sec:app-geom-general} for more details.

In contrast  for the generic spheres $\cS^4_\L$, the $p_\nu$ are  independent functions,
and \eq{MM-formula-nonloc} is replaced by  
\begin{align}
 g_{cc'}\theta^{c a} \theta^{c' b} \ &= \ \frac 14 \D^4\,(P_T^{ab} + t^{ab})\  \ =: \g^{ab}
 \label{MM-formula-generic}
\end{align}
(see  \eq{mm-id-generalS4} in appendix \ref{sec:app-geom-general})
where $t^{ab}$ has a non-vanishing radial component.
Upon averaging over the local fiber,
this defines a  scale
\begin{align}
 [\theta^{\mu 5} \theta^{\nu 5}]_{0} =: L_R^2 r^2 g^{\mu\nu}, \qquad L_R^2 = r^2 (c_n^2+N)  \geq \D^2 \ .
 \label{LR-def}
\end{align}
$L_R$ characterizes the thickness of the sphere $\cS^4_\L$ with $\L = (n_1,n_2,N)$, and
the contribution $N$ in \eq{LR-def} arises from the uncertainty\footnote{This can also be seen using e.g. 
the explicit representation of $\cS^4_N$ on $\cH = (0,N)$. Similar effects are well-known for $S^2_N$.} of $X^\mu$ 
in \eq{MM-exact}.  
In particular, this gives
\begin{align} 
 [p_\mu p_\nu]_{0} =  \frac{4L_{R}^2}{\D^4}\, g_{\mu\nu} 
 \label{PP-normaliz}
\end{align} 
and
\begin{align}
\frac{|\theta^{\mu 5}|}{|\theta^{\mu\nu}|} = \frac{L_{R}}{R}  = O(\frac{\sqrt{N+n^2}}{N}) \ .
\label{theta-P-compare}
\end{align}

\subsection{Functions versus symmetry generators}
\label{sec:kinematics}

It is  important to keep in mind the double meaning of the generators $\theta^{\mu\nu}$ and $P_\mu$:
\begin{enumerate}

\item as {\bf symmetry generators} of the  isometry group, which act on wavefunctions via the adjoint.
Then the normalization $\cM^{\mu\nu} = r^{-2} \Theta^{\mu\nu}$ is  appropriate.
 In particular  $P_\mu$ and $\cM^{\mu\nu}$ generate  the local Poincare algebra 
 for $R\to \infty$.

 \item as generators of the  algebra of  {\bf functions} on $\cO[\L]$ (along with $x^\mu$), 
 viewed as bundles over $S^4$.  In the fuzzy case, this  is replaced by $End(\cH_\L)$,
 which  describes the degrees of freedom in a field theory (or matrix model)
 on $\cS^4_\L$. This algebra is ''almost`` commutative for large $\L$.

\end{enumerate}

Consider e.g. the basic sphere $\cS^4_N$. Since the underlying space 
$\cO[\L] \cong \C P^3$ is 6-dimensional, there are locally only 6 independent 
coordinate functions. At the north pole, these are the $x^\mu$, plus 2 of the 3 selfdual $\theta^{\mu\nu}$ 
variables which parametrize the fiber $S^2$. The $\theta^{\mu\nu}$
can be viewed as  function on $\C P^3$ taking values in the SD 2-forms  $\Omega^2(S^4)$
or in the local $\msu(2)_L \subset \mso(4)$.
However, the $p^\mu$ functions vanish  in the 
semi-classical limit, as explained above. Therefore {\bf there are no  modes of the type $F_\mu(X) P^\mu$ on $\cS^4_N$},
they only exist on  the generalized spheres  $\cS^4_\L$; this will be crucial below.

Now consider the symmetry generators and their action on wavefunctions.
The quadratic Casimir of $SO(5)$ can  be written as 
\begin{align}
C^2[\mso(5)] &= \sum_{a<b\leq 5} \cM_{ab} \cM_{ab} 
 = \sum_{\mu<\nu\leq 4} \cM_{\mu\nu} \cM^{\mu\nu} +  \sum_\mu \cP_\mu \cP^\mu  \nn\\
 &= R^2\,\big(P_\mu P^\mu + \frac 1{R^2}  \cM_{\mu\nu} \cM^{\mu\nu}\big) 
\label{Casimirs}
\end{align}
using the same symbols for the abstract generators as in the  $\cS^4_N$ algebra.
Acting on scalar fields (or on fields with low spin),
the angular momentum contribution 
can be neglected\footnote{Remember that we always work in the local frame at $p$, where 
any $x^\mu$ can be dropped.}  
compared with the translational contribution,
$C^2[\mso(5)] \approx R^2\, P^\mu P_\mu$.
Comparing the formula for its eigenvalues  
\begin{align}
 C^2[\mso(5)] (n-m,2m) &= n(n+3) + m(m+1) \ 
 \label{C2-EV}
\end{align}
with the  formula for the eigenvalues of $\Box = [X^a,[X_a,.]] = C^2[\mso(6)]-C^2[\mso(5)]$ 
on $\cS^4_N$ \cite{Steinacker:2015dra} 
\begin{align}
  \Box (n-m,2m) = r^2(n(n+3) - m(m+1)), \qquad m\leq n \ ,
 \label{Box-EV}
\end{align}
it follows that
\begin{align}
  \Box \  &\approx \ r^2\, C^2[\mso(5)]| \  \approx \ - \frac{\D^4}{4} P^\mu P_\mu = -  \frac{\D^4}{4} \Box_g  , \qquad 
 \Box_g := g^{\mu\nu}\del_\mu \del_\nu
   \label{Box-Psquare}
\end{align}
for the low-spin modes  $m=0,1,2$.
A simpler way to understand this is via the semi-classical 
form of the free action  for scalar fields \cite{Steinacker:2010rh}
\begin{align}
Tr f \Box g &\sim \int - \Omega f(x) \{x^a,\{x_a,g(x)\}\}  
 = \int \Omega \{x^a,f(x)\} \{x_a,g(x)\}   \nn\\
 &=  \int \Omega  \g^{\mu\nu} \del_\mu f \del_\nu g 
 = - \int \Omega f   \g^{\mu\nu} \del_\mu\del_\nu g 
 \sim - Tr f  (\g^{\mu\nu} \del_\mu \del_\nu)  g
\end{align}
where $\Omega$ 
is the symplectic volume form, 
in  agreement with \eq{Box-Psquare}. This shows again
that $\g^{\mu\nu}$ \eq{MM-formula-metric} is the effective metric on $\cS^4_\L$.

\section{Matrix model and fluctuations  on fuzzy $\cS^4_\L$}

Now we would like to make $\cS^4_\L$ dynamical, by considering as a background in  
the Yang-Mills matrix model 
\begin{align}
 S[Y] &= \frac 1{g^2}\Tr \Big(-[Y_a,Y_b][Y^a,Y^b]\, + \mu^2 Y^a Y_a \Big) \ 
 \label{bosonic-action}
\end{align}
with a mass term as a regulator, and studying the fluctuation modes on $\cS^4_\L$.
We will later focus on the IKKT model \cite{Ishibashi:1996xs} with $D=10$.
The classical equations of motion are
\begin{align}
 (\Box + \frac 12 \mu^2) Y_a  = 0 , \qquad\qquad \Box = [Y^a,[Y_a,.]] \ .
\end{align}
We use the letter $Y$ to indicate generic configurations, while  $X$
will   indicate the fuzzy $\cS^4_\L$ background.
Although the latter is not quite a solution of these equations,
it was shown in \cite{Steinacker:2015dra} that quantum corrections
(at one loop) can stabilize the radius  of $\cS^4_\L$ for small positive $\mu^2$.
A more refined one-loop analysis will be presented in section \ref{sec:one-loop}.
Now  consider fluctuations around some fixed background $X^a$, 
\begin{align}
 Y^a =  X^a + \cA^a \ 
\end{align}
where $\cA^a \in End(\cH)$ will be the dynamical degrees of freedom.
Expanding the action expanded up to second oder in the $\cA^a$, one obtains
\begin{align}
 S[Y]  &= S[X]  + \frac{2}{g^2}\Tr \Big(2\cA^a (\Box +\frac 12\mu^2) X_a 
  +\cA_a (\Box +\frac 12\mu^2) \cA_a - 2 [\cA_a,\cA_b] [ X^a, X^b] - f^2  \Big) \ . \nn
\end{align}
dropping the linear terms (for stable backgrounds).
Hence the quadratic fluctuations $\cA^a$ are governed by the quadratic form
\begin{align}
\Tr \cA_a \Big((\Box + \frac 12 \mu^2)\d^a_b  + 2i [\Theta^{ab},. \, ] \Big) \cA_b  \ ,
\label{fluct-action-nogf}
\end{align}
where the $f^2$ term was canceled by adding a suitable 
Faddeev-Popov gauge-fixing term
(choosing the Feynman gauge  \cite{Blaschke:2011qu})
for the gauge fixing function
\begin{align}
  f = i[\cA^a,X_a] \ .
 \label{gauge-fixing-function}
\end{align}
Hence the fluctuations 
are governed by the  ``vector'' (matrix) Laplacian
\begin{align}
(D^2 \cA)_a :=
\big(\Box + \frac 12 \mu^2 - M^{(\cA)}_{rs}[\Theta^{rs},.]\big)^a_b   \cA_b \ 
\label{fluct-laplacian}
\end{align}
where 
\begin{align}
(M_{ab}^{(\cA)})^c_d &= i(\d^c_b \d_{ad} - \d^c_a \d_{bd})\, 
\end{align}
is the $SO(5)$  generator  in the vector representation.
The fluctuations $\cA^a$ entail fluctuations of the ``flux''
\begin{align}
 -i[Y^a,Y^b] = \Theta^{ab}_{(Y)}  &= \Theta^{ab}_{(X)} + \cF^{ab}, \nn\\
 \cF^{ab} &= -i[X^a,\cA^b] +i [X^b,\cA^a] -i [\cA^a,\cA^b].
\end{align}
For backgrounds given by basic noncommutative branes $\cM$, this leads to noncommutative gauge theory,
or equivalently to a theory of geometric deformations
\begin{align}
 y^a : \quad \cM \hookrightarrow \R^{10} 
\end{align}
leading to some emergent gravity on the brane which seems to be
different from general relativity \cite{Steinacker:2010rh,Steinacker:2012ra}. 
However 
on the  covariant $\cS^4_\L$ backgrounds, 
we will argue that (at least linearized)  general relativity  arises indeed from 
certain deformation modes, extended by a higher spin sector.

\subsection{Decomposition into fluctuation modes}
\label{sec:mode-decomp}

\paragraph{Global $SO(5)$ notation.}

Given some deformation $X^a + \cA^a$ of the $S^4_\L$ background, 
we want to identify the various fluctuation modes of the 5 fields $\cA^a$.
We can organize the tangential and radial fluctuations as follows, 
working mostly in the semi-classical limit 
\begin{align}
\boxed{
  \cA^a = \xi^a + \theta^{ab} \A_b  +  \frac{X^a}{R} \kappa \ .
  }
 \label{mode-expand-5D-A}
\end{align}
Here 
\begin{align}
  \xi^a &= \xi^a + \xi^{abc}\cM_{bc}  + ... , \qquad X_a \xi^a = 0  \nn\\
  \A_a &= A_b + A_{bcd} \cM^{cd} + ... , \qquad X^a \cA_a = 0\nn\\
  \kappa &= \kappa + \kappa_{ab} \cM^{ab} + ...   
 \label{mode-expand-5D-red}
\end{align}
and the functions $\xi^a,A_b,A_{abc},\k \in \cC_N(S^4) \subset End(\cH)$ play the role of tensor fields.
The expansion in $\cM$ correspond to expanding $End(\cH) = \oplus(n,2m)$  in terms of $m$.
The $\xi^a$ and the $\A_a$ are clearly tangential, and $\k$ describes the radial fluctuations.
We will only keep  tensors of rank up to $3$.
The $\A_a$ contribution is reminiscent of the standard parametrization 
in noncommutative gauge theory, and
could be interpreted as $\mmu(1)\times \mso(5)$-valued gauge field 
(or more generally as $U(\mso(5))$-valued gauge field).
Since the $X^a$ and $\cM^{ab}$ are tensor operators, there is an
$SO(5)$ action on these fields via 
\begin{align} 
 \A_a &\to  \L_a^b \, U_\L \A_b U_\L^{-1}, \qquad \quad \k \to U_\L \k U_\L^{-1} 
\end{align} 
etc., which leaves the background sphere invariant and implements the isometries on the tensor modes. 
In this sense, the theory to be elaborated will be ``covariant''. The extension to local gauge symmetries
will be discussed shortly.

Now observe that the trace sector of $A_{bcd}$ 
\begin{align}
 A_{bcd} = \frac 1{2\cR^2} \big(g_{bd}\tilde\xi _c - g_{bc}\tilde \xi_d\big) \ 
\end{align}
leads to 
\begin{align}
  \cA^a = \theta^{ab} A_{bcd}\cM^{cd}  \ = \  (P_T \tilde\xi)^a
  \label{A-B-modes}
\end{align}
using \eq{MM-formula-nonloc}, which is redundant with the $\xi^a$ modes.
Therefore we should either impose that $A_{bcd}$ is traceless, or drop the $\xi^a$ modes
(and the $\k$ modes for the generic spheres $\cS^4_\L$).
We will mostly choose the latter option. 

\paragraph{Local $SO(4)$ notation.}

To make the physical content more transparent, we will organize these fields further 
into 4D fields near some reference point $p\in S^4$ (``the north pole''). We will use 
greek indices $\mu,\nu \in\{1,...,4\}$ for tangential 
components transforming as vectors under the local $SO(4)$ around $p\in S^4$, 
and latin indices $a,b,...\in\{1,...,5\}$ for the $SO(5)$-covariant components.
In particular, the  fields will be locally expanded in powers  of the $SO(4)$-covariant generators
as in section \ref{sec:kinematics},
\begin{align}
 \Theta^{\mu\nu} &= r^2 \cM^{\mu \nu} \quad \mbox{and} \ \  P^\mu := \frac 1R \cM^{\mu 5} \ .
\end{align}
This organization gives the following modes up to the order under consideration here:
\begin{align}
\cA^\mu &= \mathbf{\xi}^\mu  \, + \frac{x^\mu} R \k \, + \theta^{\mu \nu} \A_\nu , 
\qquad \qquad  
\cA^5 = \k \ 
\label{cA-expand-4D}
\end{align}
where
\begin{align}
  \quad \A_\nu  &= A_\nu + A_{\nu\r} P^\r + A_{\nu \r\s} \cM^{\r\s} +... \nn\\
  \quad \xi^\mu &= \xi^\mu(x)  + \xi^{\mu \nu}(x)P_{\nu} + \xi^{\mu \nu\r}(x)\cM_{\nu\r} +... \nn\\
 \quad\k &= \k + \k_{\mu} P^\mu  + \k_{\mu\nu} \cM^{\mu\nu}  + ... 
  \label{4D-modes-decomp}
\end{align}
where both $A_{\nu \r\s}$ and $A_{\nu\r} = A_{\nu\r 5}$ arise form the 5D fields $A_{abc}$.
We separate  $A_{\mu\nu}$ into symmetric and antisymmetric (AS) parts, 
\begin{align}
 A_{\nu \r} &= \frac 12(h_{\nu \r} + a_{\nu \r}) , \qquad h_{\nu \r} = h_{\r\nu}, 
 \qquad a_{\nu \r} = -a_{\r\nu} \ . 
\end{align}
 As discussed above, we can 
 absorb $\xi^\mu$ in 
 \begin{align}
 \tilde A_{\nu \r \s} = A_{\nu \r \s} + A_{\nu \r \s}[\xi], \qquad 
  A_{\nu \r \s}[\xi] = \frac 1{\cR^2} (P_{SD})_{\r\s}^{\r'\s'}g_{\nu \s'}\xi _{\r'} \ .
  \label{A-trace-xi}
 \end{align}
 Here $P_{SD}$ is the projector on the SD antisymmetric component.
%
Similarly, the $\A_5$ modes can absorb the radial $\k$ modes for the generic spheres, but should be dropped for 
$S^4_N$ since the corresponding fluctuations $\cA^\mu \sim P^{\mu} A_5$ vanish.

We will see  that the  $A_\mu$ describes a $U(1)$ gauge field and 
$h_{\mu\nu}$ determines the metric fluctuations, while $a_{\mu\nu}$ does not seem to play a significant role. 
$A_{\mu\nu\r}$ is  part of the gravitational sector.
As discussed in the previous section, the $h_{\mu\nu}$ modes only exist on the generalized spheres $\cS^4_\L$,
while they vanish on the basic $\cS^4_N$ due to the tangential constraint \eq{theta-P-exact}. 

\begin{figure}
\begin{center}
 \includegraphics[width=0.4\textwidth]{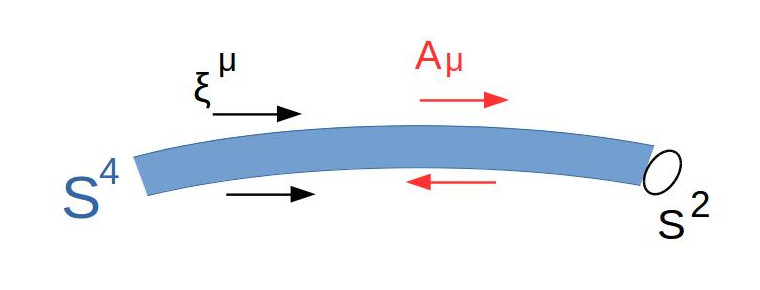}
 \end{center}
 \caption{Schematic local picture of the  deformation modes $A_\mu$ and $\xi^\mu$.}
 \label{fig:deformations}
\end{figure}

It is important to keep in mind that (apart from the  $\xi^\mu$ and the  $\k$ deformations)
 these deformation modes are ``internal''
degrees of freedom, whose averages $[.]_{0}$ over the local fiber vanishes.
Some of these deformations are sketched in figure \ref{fig:deformations}.
The only modes which change the embedding of  $S^4$  in target space are the radial modes $\k$.
The organization \eq{4D-modes-decomp} is quite general
and applies also to other covariant quantum spaces, even  with Lorentzian signature. 
The full expansion into higher spin modes is obtained by allowing
the $\A_\mu, \xi^\mu $ and $\k$  fields to
take values in the universal enveloping algebra of $\mso(5)$ or $I\mso(4)$.

\subsection{Gauge transformation}

Consider gauge transformations
\begin{align}
 Y^a  \ &\to \ Y^a + i [\L,Y^a] 
\end{align}
with some gauge parameter $\L\in End(\cH)$. 
For fluctuations on a background
$Y^a =  X^a + \cA^a$, this leads to the inhomogeneous transformation
\begin{align}
  \d \cA^a &= i [\L,X^a] + i [\L,\cA^a] \ .
\end{align}
We can expand the gauge parameter in $SO(5)$ generators as  
\begin{align}
  \L &= \L_0 +  \frac 12 \L_{ab} \cM^{ab} + ...
\end{align}
where $\L_0, \ \L_{ab} \in \cC_N(S^4)$.
Clearly $\L_{ab}\cM^{ab}$ generates an $x$-dependent $SO(5)$ transformation, and 
\begin{align}
 \A_a = A_a(x) + A_{abc}(x) \cM^{bc}
\end{align}
transforms as (noncommutative) $SO(5) \times U(1)$-valued gauge field.

\paragraph{Local $SO(4)$ rotations \& diffeomorphisms.}

It is interesting to work out the explicit form of these transformations in the local
4D parametrization \eq{4D-modes-decomp}. We expand
\begin{align}
  \L &= \L_0 + v_\mu P^{\mu} + \frac 12 \L_{\mu\nu} \cM^{\mu\nu} + ... \ .
  \label{gauge-parameter}
\end{align}
and define the individual transformations
\begin{align}
 \d_{\L_0} &:= i[\L_{0},.] ,  \nn\\
 \d_v  &:= i[ v_\r P^\r,.] , \nn\\
 \d_\L &:= \frac i2[\L_{\r\s}\cM^{\r\s},.] .
\end{align}
In the semi-classical limit, we can replace the commutators by Poisson brackets, and
 \begin{align}
  \d_{\L_0} X^\mu  &= \theta^{\mu\nu} \del_\nu \L_0  \nn\\
  \d_v X^\mu  &= - v^\mu + \theta^{\mu\nu} (\del_\nu v_\r ) P^\r \nn\\
 \d_v \phi &= \ i[ v_\r P^\r,\phi] 
  \qquad \quad \ \sim \  - v^\r \del_\r \phi  \  + \theta^{\mu\nu}(\del_\mu \phi)(\del_\nu v_\rho P^\r)    \nn\\
\d_\L X^\mu &= \frac 12 i[\L_{\r\s}\cM^{\r\s},X^\mu]  \ \sim \ \frac 12 \theta^{\mu\nu}\del_\nu \L_{\r\s} \cM^{\r\s} \nn\\
  \d_\L \phi &= \frac 12 i[\L_{\r\s}\cM^{\r\s},\phi] 
   \ \ \  \sim \   O(\theta \del\phi\del\L)  
 \label{gauge-trafo-v-L}
 \end{align}
 where $\phi = \phi(X)$ indicates some scalar field, 
  and $\theta^{\mu\nu}$ is the undeformed Poisson tensor.
  Here we recalled that $x^\mu = 0 = \d_\L x^\mu$ at $p$.
 Restricted to the lowest degree in $\theta$,
 the $\d_v$ clearly acts as a diffeomorphisms on scalar fields $\phi(x)$, and
 $\d_\L$ leads to local $SO(4)$ rotations of tensors 
 (which vanishes for scalar functions at the north pole). 
 Applying this to the background $X+\cA$, we can read off the transformations of the 
 tangential and radial perturbations 
\begin{align}
 \d\cA^\mu &= \theta^{\mu\nu} \del_\nu (\L_0 +  v_\r P^{\r} + \frac 12\L_{\s\r} \cM^{\s\r} ) -  v^\mu
 + \d_v\cA^\mu + \d_\L\cA^\mu  + \d_{\L_0}\cA^\mu \nn\\
  \d\cA^5 &= \d_v\k + \d_\L\k + \d_{\L_0}\k
  \end{align}
  where
 \begin{align}
 \d_v(\theta^{\mu \nu} \A_\nu) 
 &=  i[v_\rho P^\r,\theta^{\mu \nu} ] \A_\nu +  \theta^{\mu \nu} \d_v \A_\nu    \nn\\
   &= v_\rho r^2 (-g^{\r\mu}P^\nu +  g^{\r\nu}P^\mu) \A_\nu +  \theta^{\mu \nu} \d_v \A_\nu \nn\\
\d_\L(\theta^{\mu \nu} \A_\nu) 
   &= i\frac 12[ \L_{\s\r}\cM^{\r\s},\theta^{\mu \nu} ] \A_\nu +  \theta^{\mu \nu}\d_\L \A_\nu    \nn\\
   &=  -\L_{\s\r}(g^{\r\mu}\theta^{\s \nu} - g^{\r\nu}\theta^{\s \mu}) \A_\nu +  \theta^{\mu \nu}\d_\L \A_\nu  \nn\\
   &= (\L \cdot \theta \A)^\mu  + \theta^{\mu\nu} (\L\cdot \A)_\nu  \ .
 \end{align}
Here we denote the local rotation of  $\A_\mu$ by $\L\in \mso(4)$ with
\begin{align}
(\L\cdot \A)_\mu := -\L_{\mu\r}g^{\r\nu} \A_\nu + ...
\end{align}
which extends to all the tensor legs of $\A_\mu$ in the expansion \eq{4D-modes-decomp}.
Dropping  contributions to the higher $\xi^\mu$ modes which we don't keep track of, 
we obtain the following linearized gauge transformations for the 4D fields
 \begin{align}  
 \d \A_\mu &= \del_\mu \big(\L_0 +  v_\r P^{\r} + \frac 12 \L_{\s\r} \cM^{\s\r}\big)  -  v^\rho \del_\rho \A_\mu  
  + (\L\cdot \A)_\mu  - v_\rho \theta^{\r\mu} \nn\\
 \d\k \ &= -  v^\rho \del_\rho \k  \nn\\  
 \d\xi^\mu &= -  v^\mu  \ .
 \label{gaugetrafo-fluct}
\end{align}
Separating $\A_\mu$ into the tensor components, this gives
\begin{align}
 \d A_\mu &=  \del_\mu \L_0 -  v^\rho \del_\rho A_{\mu}  + (\L\cdot A)_\mu  \nn\\
 \d a_{\mu\nu} &= (\del_\mu v_\rho - \del_\r v_\mu) -  v^\rho \del_\rho a_{\mu\nu}  + (\L\cdot a)_{\mu\nu} \nn\\
 \d h_{\mu\nu} &=  (\del_\mu v_\rho + \del_\r v_\mu) -  v^\rho \del_\rho h_{\mu\nu}  + (\L\cdot h)_{\mu\nu} \nn\\
 \d A_{\mu\r\s}  &= \frac 12 \del_\mu \L_{\s\r}(x) - v^\rho \del_\rho A_{\mu\r\s} + (\L\cdot A)_{\mu\r\s} \ .
 \label{gaugetrafo-fields}
\end{align}
These can be understood as 
local $SO(4)$ rotations generated by $\L_{\mu\nu}(x)$, $U(1)$ gauge transformations generated by $\L_0(x)$,
and infinitesimal diffeomorphisms generated by $-v^\rho \del_\rho$. 
The $A_{\mu\r\s}$ transforms like a $SO(4)$ gauge field.
The inhomogeneous transformation of $h_{\mu\nu}$ under  diffeomorphisms can be understood  by
anticipating that it plays the role of a linearized metric fluctuation
$g_{\mu\nu} -  h_{\mu\nu}$;
its transformation by $v$  then gives
\begin{align}
\d (g_{\mu\nu} -  h_{\mu\nu})
   &= g_{\mu\nu} -  h_{\mu\nu}
   -\big(\del_\mu v_\rho + \del_\r v_\mu + v^\rho \del_\rho (g_{\mu\nu} -  h_{\mu\nu}) ) ,
\end{align}
which is  the transformation of the metric tensor $g_{\mu\nu} -  h_{\mu\nu}$
under an infinitesimal diffeomorphism $- v^\r \del_\r$.

\paragraph{Higher-order gauge transformations.}

The gauge transformations considered in \eq{gauge-parameter} are only the lowest in a whole tower.
Consider e.g. the transformations generated by
\begin{align}
  \L &= \L_{\a\b} P^{\a}  P^\b \ ,
  \label{gauge-parameter-PP}
\end{align}
which leads to 
\begin{align}
 [X^\mu,\L] = -i\L_{\a\b} (g^{\mu\a} P^\b + g^{\mu\b} P^\a) + [X^\mu,\L_{\a\b}] P^{\a}  P^\b \ .
\end{align}
This allows to gauge away the symmetric $\xi^{\mu\nu}$ modes. 
In contrast, one cannot gauge away the $h_{\mu\nu}$  modes.
From a geometric point of view, the pure gauge modes correspond to Hamiltonian vector fields on $\C P^3$, and
a systematic analysis  is postponed for future work.

\paragraph{Gauge fixing.}

The gauge fixing was achieved by adding the Faddeev-Popov (or BRST)
gauge fixing term $-f^2$ to the action, such that the explicit
$f^2$ term in \eq{fluct-action-nogf} is canceled. This ensures that the propagator 
is well-defined. The corresponding gauge fixing condition $0 =  i[X_a,\cA^a] $ is 
accordingly not a ``hard constraint", but simply selects the 
physical Hilbert space or configuration space without redundancies.

Now consider the gauge fixing condition 
\begin{align}
 0 =  i[X_a,\cA^a] = i[X_5,\cA^5] +i[X_\mu,  \theta^{\mu\nu} \A_\nu] \ .
 \label{gaugefixing-decomp}
\end{align}
The radial contribution from $\k$ is
\begin{align}
  i[X^a,\frac {X_a}{R}\k] &\sim - \{x^5,\k\} = \theta^{\mu 5}\del_\mu \k 
 =  \frac 12 r \D^2  P^\mu \del_\mu \k \ ,
\end{align}
%
thus the gauge fixing condition is 
\begin{align}
 0 &\sim \ - \frac 14 \D^4 g^{\nu\mu}\del_{\mu} \A_\nu +  \frac 12 r \D^2  P^\mu \del_\mu \k
  + \theta^{\mu\nu}A_{\nu \mu}  \ .
   \label{gauge-fixing-A}
\end{align}
Separating the components,
this leads to  
\begin{align}
 0 &= \del^\mu A_{\mu}\nn\\
 0 &= \frac {\D^2}{2} \del^\mu A_{\mu\nu}  -  r \del_\nu \k   \nn\\
 0 &=  \theta^{\mu\nu}\big(\frac 12 a_{\nu \mu} 
 - R^2 \del^\rho \tilde A_{\rho\mu\nu} \big)  
 \label{gaugefix-components}
\end{align}
For $a_{\mu\nu} = 0$, these reduce to the Lorentz gauge condition for $A_\mu$ and $A_{\rho\mu\nu}$ 
while the second condition reduces to 
$\del^\mu h_{\mu\nu} = \frac 2R \del_\nu \k$. 

\section{Geometry: metric and vielbein}
\label{sec:geometry}

\paragraph{Undeformed background.}

Consider  some scalar field $\phi = \phi(X)$. The adjoint action of the basic matrices 
$[X^a,.]$ defines a derivative operator on $\phi$,
\begin{align}
 D^a \phi &:= -i[X^a, \phi] \ \sim  e^{a \mu} \del_\mu\phi \ . 
\end{align}
where
\begin{align}
 e^{a \mu}= \theta^{a\mu},   \qquad \quad e^a = e^{a \mu}\del_\mu 
 \label{vielbein}
\end{align}
plays the role of a vielbein or frame. Using  \eq{theta-P-compare}, 
we see that the tangential 
vielbeins 
\begin{align}
 e^\a \sim \theta^{\a\mu}\del_\mu, \qquad \a=1,...,4
\end{align}
play the dominant role, 
while the transversal component
 $e^5 \sim  \theta^{5\mu}\del_\mu$
only contributes a small multiplicative factor $\frac{L_R^2}{R^2}$ to $\g^{\mu\nu}$ via \eq{PP-normaliz}.
Recalling the discussion in section \ref{sec:kinematics}, this vielbein arises from
the bundle of (selfdual) 2-tensors\footnote{Note that $\theta^{\mu\nu}$ is no longer self-dual for the generalized spheres $\cS^4_N$.
Then the treatment in this and the following sections should be generalized accordingly. However this will not 
lead to significant changes, and we stick to the self-dual case here for simplicity.}  $\theta^{\mu\nu}$, 
which transform in the $(1,0)$ under $SO(4)$ along the internal fiber $S^2$.
Hence $e^{\a \mu}$ is not a fixed frame on $S^4$, but it is averaged out over the fiber, $[e^{\e\nu}]_{0} = 0$.
We can now rewrite the formula for the metric \eq{MM-formula-metric} on $\cS^4_\L$ in a more suggestive way as follows
\begin{align}
 \bar\g^{\mu\nu} = g_{ab}\, \theta^{a\mu}\theta^{b\nu} = g_{ab} e^{a \mu}\, e^{b \nu}
 \label{G-theta}
\end{align}
This defines a fixed, well-defined  metric on $S^4$ which is constant along the fiber $S^2$,
\begin{align}
  [\bar\g^{\mu\nu}]_{0} = \bar\g^{\mu\nu}  = \frac{\D^4}{4} g^{\mu\nu} .
\end{align}
This is the key property which allows  to reconcile covariance with noncommutativity.
For generic $\cS^4_\L$ with large $L_R$, the $\bar\g^{\mu\nu}$ is  replaced by the 5-dimensional $\bar\g^{ab}$
as in \eq{MM-formula-generic}.

Now consider  general fields $\phi \in End(\cH)$,
decomposed into a tower of higher spin (tensor) 
fields on $S^4$ as in \eq{NC-modes}.
The adjoint action $[X^a,.]$ still defines a derivative operator $\phi$,
which however contains  non-derivative terms 
 which arise from  commutators of the $X^\a$ with the $P^\mu$ generators in 
the expansion of $\phi$. 
E.g. for   $\phi  = \phi + \phi_\mu P^\mu + \phi_{\mu\nu} \cM^{\mu\nu}$, we have
 \begin{align}
  D^\a \phi  &= -i[X^\a,\phi + \phi_\mu P^\mu + \phi_{\mu\nu} \cM^{\mu\nu}]  
   \sim  e^{\a\r} \del_\r\phi -   \phi^{\a} \ . 
 \label{D-phi-S4}
 \end{align}
This phenomenon will play a crucial role below.
Nevertheless, the metric in the kinetic term for arbitrary fields is always obtained from the 
leading derivative contributions
\begin{align}
 -[X^a,\phi][X_a,\phi] \sim  \bar\g^{\mu\nu}\del_\mu\phi \del_\nu\phi + \dots \ .
 \label{bracket-metric}
\end{align}

\paragraph{Deformed background.}

Now we include the  fluctuations $Y^a = X^a + \cA^a$.
Since the kinetic term for (bosonic) fields always arise from contracted commutators 
\begin{align}
 -[Y^a,\phi][Y_a,\phi] \ &= \ D^a \phi  D_a \phi \sim  \g^{\mu\nu} \del_\mu \phi  \del_{\nu} \phi + ... , 
  \qquad D^a := -i[Y^a,.] 
 \label{kinetic-term}
\end{align}
we can read off the effective  metric in the perturbed matrix model
(up to a possible conformal factor, see below) 
\begin{align} \boxed{
 \  \g^{\mu\nu} \sim  \{Y^a,X^\mu\}\{Y_a,X^\nu\} \  =  \ D^a x^\mu  D_a x^\nu \ .
 }
 \label{metric-vielbein-gen}
\end{align}
and similarly for the 5D case with $\g^{ab}$.
This can be expressed in terms of an 
(over-complete) frame
\begin{align} \boxed{
 \  e^a [\cA] =  e^{a\mu}[\cA]\del_\mu  \ \  = D^a\  \ 
   }
 \label{vielbein-e}
\end{align}
cf. \cite{Steinacker:2010rh,Steinacker:2012ct}. 
Again the tangential contributions $e^\a[\cA], \ \a=1,...,4$ will provide the leading contribution.
Recall the explicit form of tangential fluctuations
$\cA^\a = \frac{x^\a}{R} \k  + \theta^{\a\mu} \A_\mu$ with
 \begin{align}
  \A_\mu  &= A_\mu + A_{\mu \nu} P^{\nu} + \tilde A_{\mu \r\s} \cM^{\r\s}\ ,
 \end{align}  
Observe first
\begin{align}
 -i[\A_\mu,\phi]  &\sim (A_{\mu\b}g^{\b\nu} + \theta^{\b\nu}\del_{\b} \A_\mu)\del_\nu\phi  
\end{align}
where the non-derivative $A_{\mu\nu}$ term arises from the explicit $P$ modes in $\A_\mu$,
similar as in \eq{D-phi-S4}. 
Using these expressions and dropping as usual $[\cM^{\a\b},\phi] = O(\frac{x}{R}\del\phi)$ in the local frame,
we obtain\footnote{We include also the term $\phi^\mu P_\mu$ in the expansion of $\phi\in End(\cH)$,
which is needed for the field strength where $\phi \to \cA$.
The term $\del_{\b} A_{\mu\r}P^{\r}$ is dropped since it would drop out upon averaging over
$S^2$ in \eq{del-gamma-average}.}
\begin{align}
 D^\a \phi &\sim -i[Y^\a,\phi] = -i[X^\a(1+\frac{\k}{R}) + \cA^\a,\phi]  \nn\\
  &\sim 
   e^{\a\nu}[\cA]\del_\nu \phi - \phi^\a \  \ .
 \label{D-phi-general}
\end{align}
 where the tangential vielbein $e^\a[\cA] = e^{\a\nu}[\cA] \del_\nu$ is obtained as
\begin{align}
 e^{\a\nu}[\cA] &= \{Y^\a,X^\nu\} 
  = \theta^{\a\mu}\big(\d^\nu_{\mu}(1+ \frac{\k}R) +  A_{\mu\b} g^{\b\nu}  
  + \theta^{\b\nu} (\del_{\b} A_{\mu} + \del_{\b} \tilde A_{\mu\r\s}\cM^{\r\s})\big)\,\ .  
 \label{vielbein-A}
\end{align}
The transversal vielbein 
\begin{align}
 e^{5\nu}[\cA] &= \{Y^5,X^\nu\}
 =  \theta^{5\nu} + \theta^{\mu\nu}\del_\mu\kappa  
\end{align}
does not contribute to the linearized metric perturbations and can be dropped.
Hence the tangential $e^\a, \ \a=1,...,4$ play the role of the effective vielbein. 
Using these results,  the metric  on a deformed  $\cS^4_\L$  background 
including linearized perturbations is 
\begin{align} 
 \  \g^{\mu\nu}  \ =  \{Y^\a,X^\mu\}\{Y_\a,X^\nu\} \
 = g_{\a\b} e^{\a\mu} e^{\b\nu} =  \bar\g^{\mu\nu} + \d \g^{\mu\nu} 
 \label{metric-vielbein-4}
\end{align}
where the linearized metric fluctuations are given by
\begin{align}
 \d \g^{\mu\nu} 
  &= \frac{\D^4}{4} \big(h^{\mu\nu} +  \frac{2\k}R g^{\mu\nu} \big)
   + \frac{\D^4}{4}\Big( g^{\mu\nu'}  \theta^{\b\nu} (\del_{\b} A_{\nu'} + \del_{\b} \tilde A_{\nu'\r\s}\cM^{\r\s}) 
   + (\mu \leftrightarrow \nu) \Big) 
 \label{graviton-linear}
\end{align}
using \eq{MM-formula-metric},
 always raising and lowering indices with $g^{\mu\nu}$.
 Note that the anti-symmetric contributions $a_{\mu\nu}$ drop out. 
 After averaging over the fiber $S^2$ using
 \eq{M-correlators-2sphere}, the 
 contribution from the $U(1)$ gauge field $A_{\nu}$ also drops out since $[\theta^{\b\nu}]_{0} =  0$, and we obtain
 \begin{align}
 \boxed{
   \left[{\d \g}^{\mu\nu}\right]_{0} 
   =  \frac{\D^4}{4}\Big(h^{\mu\nu} + k^{\mu\nu}  + \frac{2\k}R g^{\mu\nu} \Big)\ =: \frac{\D^4}{4}\tilde h^{\mu\nu} .
   }
 \label{del-gamma-average}  
 \end{align}
 The contribution from the rank 3 tensor $\tilde A_{\mu \r\nu}$  is
 \begin{align} 
 k_{\mu\nu} &:= g^{\mu\nu'}[\theta^{\b\nu} \cM^{\r\s}]_{0} \del_{\b} \tilde A_{\nu'\r\s} 
      + (\mu \leftrightarrow \nu) =  \frac {4R^2}3 (\del^\r \tilde A_{\mu \r\nu}  + \del^\r \tilde A_{\nu \r\mu})
 \label{kmunu-def-1}
\end{align}
 using  $\frac{\D^4}{4} = r^2 R^2$ as well as the self-duality of
$\tilde A_{\mu\r\s}$  in the last indices. 
Note that $k_{\mu\nu}$ transforms as 
 \begin{align}
  k_{\mu\nu} \to k_{\mu\nu} + \del_\mu v_{\nu} + \del_\nu v_{\mu} , 
  \qquad v_\nu = \frac {4R^2}3 \del^\rho \L_{\nu\r}
 \end{align}
 under local $SO(4)$ gauge transformations, and
 the gauge condition \eq{gaugefix-components} for $\tilde A_{\mu \r\nu}$ implies 
\begin{align}
 \del^\mu k_{\mu\nu} = 0  \qquad \mbox{if} \quad
 a_{\mu\nu} = 0 \ \ .
\end{align}
%
%
%
%

\subsection{Thick spheres, extra dimensions and dimensional reduction} 
 \label{sec:dimred}
 
 Since the underlying space $\cO[\L]$ is higher-dimensional, there are excitation modes
 in  extra dimensions. Most of them give rise to higher spin modes, as discussed above. 
 However for the generalized spheres $\cS^4_\L$, there are also extra scalar modes, corresponding 
 to the  $SO(5)$ Casimir $\cR^2$.
 For long-range gravity, we will  actually need very ``thick`` spheres $\cS^4_\L$,
 with large $n \gg \sqrt{N}$ \eq{cn-estimate-gravity}. 
 In order to obtain nevertheless 4-dimensional physics, we  need to assume dimensional reduction,
 i.e. all wavefunctions are constant in these extra directions. More precisely, we assume  that only the lowest
 non-trivial modes \eq{4D-modes-decomp} in these directions are significant, essentially leading to gravity.

 There are several possible justifications for this dimensional reduction. 
 First, since (most of) these internal excitation modes lead to higher spin fields, their interactions will be suppressed 
 by a large mass scale, which should arise from the scale of noncommutativity\footnote{This is in contrast to standard higher spin theories,
 where the only available scale is the IR scale. I thank K. Mkrtchyan for useful discussions on this point.}.
 Furthermore, quantum effects will certainly induce some mass gaps, in particular for the radial direction, cf.  \cite{Steinacker:2015dra}.
  Finally  there is another  mechanism to  give masses to  internal modes,
 by explicitly embedding   extra dimensions in the matrix model,
 along the lines of  \cite{Aschieri:2006uw,Steinacker:2014lma,Medina:2002pc}.
 This can easily lead to  a large mass gap\footnote{The gravitational modes are protected 
 from acquiring a mass by gauge invariance.} in the  fuzzy extra 
 dimensions, and at the same time lead to an interesting low-energy gauge theory.
 For example, the 6 transversal matrices in the IKKT model could be identified 
 with the generators of squashed $\C P^2$ \cite{Steinacker:2014lma,Steinacker:2015mia}, since
 $\cS^4_\L$ is a $\C P^2$-bundle over $\cS^4_N$, see appendix \ref{sec:app-geom-general}.
 This will be studied  elsewhere. Here we will simply assume
 dimensional reduction to 4-dimensions, and 
 elaborate the resulting  4-dimensional gravity.
 In any case, a significant asymmetry $0 \ll n \ll N$ of $\L=(n,0,N)$ is presumably essential to justify this 
  dimensional reduction.

\subsection{Effective metric and scalar fields}
 \label{sec:coupl-scalar-matter}

To properly identify the effective metric, consider scalar fields propagating on the deformed $\cS^4_\L$ background
in more detail.
The kinetic term for  a (transversal) scalar field is 
\begin{align} 
 S[\phi] &= - \frac 2{g^2} \Tr [Y^a,\phi][Y_a,\phi] 
  \ \sim \ \frac {\dim \cH}{\rm Vol(\cM^4)}\frac{2}{g^2} \int_\cM d^4 x\, 
  \g^{\mu\nu}\del_\mu \phi \del_\nu \phi \ 
\end{align}
To be specific, we use the Riemannian measure in target space.
We can cast this into a covariant-looking form (cf. \cite{Steinacker:2010rh})
\begin{align} 
 S[\phi] &\ \sim \ \frac {\dim \cH}{\rm Vol(\cM^4)}\frac{\D^4}{2 g^2} \int_\cM d^4 x\, \sqrt{|G_{\mu\nu}|}\,
  G^{\mu\nu}\del_\mu \phi \del_\nu \phi \nn\\
  &= \frac 12 \int_\cM d^4 x\,  \sqrt{|G_{\mu\nu}|}\,
  G^{\mu\nu} \del_\mu \varphi \del_\nu \varphi \
\end{align}
in terms of the {\em effective metric}
\begin{align}
\boxed{
 \ G^{\mu\nu} = \a \frac{4}{\D^4}  \g^{\mu\nu},\qquad \a = \sqrt{\frac{\D^4}{4} |\g^{-1}_{\, \mu\nu}| } \ 
 = 1-\frac 12 \tilde h + ...\ 
 }
 \label{eff-metric-G}
\end{align}
 which is a dimensionless Weyl rescaling of $\g^{\mu\nu}$, and a field $\varphi$ which has dimension mass via
 \begin{align}
  \phi = \frac{\D^2\, g_{\rm YM}}{2\sqrt{2}\,}\, \varphi \ .
  \label{phi-varphi}
 \end{align}
 Here $g_{\rm YM}$ is defined in \eq{g-YM-def}, and
the corresponding linearized metric fluctuation is obtained from \eq{del-gamma-average} 
\begin{align}
 G^{\mu\nu} &= g^{\mu\nu} + H^{\mu\nu}, \qquad 
 H^{\mu\nu} =  \tilde h^{\mu\nu}  - \frac 12 g^{\mu\nu} \tilde h \ .
 \label{H-def}
\end{align}
where $\tilde h_{\mu\nu} = h_{\mu\nu} + k_{\mu\nu}$.
Then the Lorentz-gauge condition $\del_\mu \tilde h^{\mu\nu} = 0$
translates into the de Donder gauge for $H_{\mu\nu}$,
\begin{align}
  \del^\mu H_{\mu\nu} - \frac 12 \del_\nu H \ = 0 \  .
  \label{dedonder}
\end{align}
We consider $H_{\mu\nu}$ as a tensor field here, rising and lowering indices with $g_{\mu\nu}$.
 Then the linearized coupling of $\tilde h_{\mu\nu}$ to matter is given by
 \begin{align}
  \d_h  S[\phi] &=  \frac12 \int_\cM d^4 x\, H^{\mu\nu} T_{\mu\nu}[\varphi] 
    = \frac 12 \int_\cM d^4 x\, \tilde h^{\mu\nu}
   \ (T_{\mu\nu}[\varphi]-\frac 12 g_{\mu\nu} T)  
 \label{scalar-coupling-linearized}
 \end{align}
  where 
\begin{align}
 T_{\mu\nu}[\varphi] =  \del_\mu\varphi \del_\nu\varphi 
  - \frac 12 g_{\mu\nu} (g^{\r\s}\del_\r\varphi \del_\s\varphi)  , \qquad T = -g^{\mu\nu}\del_\mu\varphi \del_\nu\varphi 
\end{align}
is the energy-momentum tensor of $\varphi$, which satisfies 
 $\del^\mu T_{\mu\nu} = 0$. 

\subsection{Flux and field strength}
\label{sec:flux}

Now consider the perturbation of the ``flux''  $[Y^\mu,Y^\nu] \sim i\Theta^{\mu\nu}_{(Y)}$ given by
\begin{align} 
 \Theta^{\mu\nu} &= \theta^{\mu\nu} + \cF^{\mu\nu} 
  = \tilde\theta^{\mu\nu}[A]  + \theta^{\mu\mu'} \theta^{\nu \nu'} F_{\mu'\nu'}[\A]   \nn\\[1ex]
  \tilde\theta^{\mu\nu} &:=  \theta^{\mu\nu} + \theta^{\mu \mu'}  A_{\mu'\nu'} g^{ \nu'\nu} - \theta^{\nu \nu'} A_{\nu'\mu'}g^{\mu' \mu}  \nn\\
   &= (\d^\mu_{\mu'} +  A_{\mu'\r}g^{\mu\r})\theta^{\mu'\nu'}(\d^\nu_{\nu'} + A_{\nu'\r} g^{ \r\nu} )
  \label{flux-perturbed}
\end{align}
to linearized order.
Since the  $A$ terms enter through one factor of $\theta^{\mu\mu'}$, 
 they are naturally viewed as geometric deformation of the background 
 $\theta^{\mu\nu}\to\tilde\theta^{\mu\nu}$, which 
plays the role of the  Poisson tensor in the deformed $y^\mu$ coordinates.
In contrast, the field strength
 \begin{align}
   F_{\mu\nu}[\A]  &= \del_{\mu} \A_{\nu} -  \del_{\nu} \A_{\mu} \ -i[A_\mu,A_\nu] \nn\\
   &=  F_{\mu\nu} +  R_{\mu\nu}   + T_{\mu\nu} 
 \end{align}
enters via 2 factors of $\theta$, and decomposes into the $U(1)\times I\mso(4)$-valued components
 \begin{align}
   F_{\mu\nu} &= \del_{\mu} A_{\nu} - \del_{\nu} A_{\mu}   \nn\\
   R_{\mu\nu} &= \del_{\mu} \omega_{\nu} - \del_{\nu}  \omega_{\mu}  -i [\omega_\mu,\omega_\nu], 
   \qquad \qquad\qquad \  \omega_\mu = A_{\mu\a\b} \cM^{\a\b}  \nn\\ 
   T_{\mu\nu} &= \del_{\mu} \a_{\nu} -   \del_{\nu}  \a_{\mu}  -i ([\omega_\mu,\a_\nu] - [\omega_\nu,\a_\mu]), 
    \qquad \a_\mu = A_{\mu\a} P^{\a} \ .
    \label{F-R-T-explicit}
 \end{align}
Clearly $F_{\mu\nu}$ is the $U(1)$ field strength of $A_\mu$, and
$R_{\mu\nu}$ is the curvature of $A_{\mu\a\b}\cM^{\a\b}$ viewed as $\mso(4)$ connection.
Furthermore, $T_{\mu\nu}$ can be related to the linearized spin connection.

\paragraph{Stack of branes and nonabelian gauge theory.}

As usual, $SU(k)$ gauge fields can be obtained by considering a stack of coincident 
branes,  expanding the fluctuations as $A_\mu = A_\mu \one + A_{\mu,a} \l^a$ 
in terms of $\msu(k)$ generators $\l^a$. 
For the generic $\cS^4_\L$ spheres there is no need to do this by hand, since they 
can be interpreted as bundles over $\cS^4_N$ with fibers being fuzzy coadjoint orbits 
of $SU(3)$ (see Appendix \ref{sec:app-geom-general} for more details, and \cite{Grosse:2004wm}
for an explicit example in a simplified setting). 
This means that some non-trivial gauge theory will arise automatically, whose structure is similar 
to the squashed brane configurations  in \cite{Steinacker:2015mia,Steinacker:2014lma}, 
which in turn are quite close to the standard model. 
It is very remarkable that  the  $\cS^4_\L$ spheres seem to provide the right ingredients for 
both gravity and  particle physics.

For such nonabelian gauge fields arising on $\cS^4_\L$,
the $\mmu(k)$-valued fluctuations  should be expanded 
in terms of\footnote{This expansion should be applied also in \eq{cA-expand-4D} 
to go beyond the present linearized approximation.}
\begin{align}
 \cA^\a =  \tilde e^{\a\nu}A_\nu
 \label{cA-expand-e}
\end{align}
rather than $\cA^\a = \theta^{\a\nu}A_\nu$.
Here
\begin{align}
  \tilde e^{\a\mu}[A] &= \sqrt{\a}\frac{2}{\D^2}\, e^{\a\mu}[A] \ 
  \label{vielbein-rescaled}
\end{align} 
is the conformally rescaled dimensionless vielbein corresponding to the 
effective metric $G^{\mu\nu}$ \eq{eff-metric-G}.
Then the correct coupling to the metric  is recovered (cf. \cite{Steinacker:2010rh}),
\begin{align}
 S[F] &= \frac 1{g^2} tr [Y^\mu,Y^\nu][Y_\mu,Y_\nu] 
 \ \sim \ \frac 1{4 g_{\rm YM}^2} \int_\cM d^4 x\, \sqrt{|G|}\,
  G^{\mu\nu}G^{\mu'\nu'} F_{\mu\mu'}  F_{\nu\nu'} \ .
\end{align}
where the Yang-Mills coupling constant is defined as\footnote{Note that the matrix model coupling $g^2$ has 
dimension $L^4$. For nonabelian gauge fields,  an extra factor may arise from the 
number of branes.}
\begin{align}
 \frac 1{4 g_{\rm YM}^2} = \frac {\dim \cH}{\rm Vol(\cM^4)}\frac{\D^8}{16g^2}
 \label{g-YM-def}
\end{align}
Noting that the conformal factor drops out in the Yang-Mills action,
the linearized coupling to the metric perturbation $h_{\mu\nu}$ gives 
\begin{align}
 \d_h S[F]  &= \frac 12 \int_\cM d^4 x\, h^{\mu\nu} \, T_{\mu\nu}[F]
   \label{flux-coupling-linearized}
\end{align}
where 
\begin{align}
 T_{\mu\nu}[F] =  \frac 1{g_{\rm YM}^2} \big(F_{\mu\mu'}  F_{\nu\nu'} G^{\mu'\nu'}  - \frac 14 G^{\mu\nu}(F F)\big)
\end{align}
is the energy momentum tensor of the gauge fields.

\section{Gravity}

\subsection{Classical action and equations of motion}
\label{sec:eom}

In order to derive the equations of motion for the gravitational $A_{\mu\nu}, A_{\mu\r\s}$ and $\k$ modes,
we evaluate the semi-classical action up to quadratic order.
The quadratic fluctuations are governed by the  ``vector'' (matrix) Laplacian \eq{fluct-laplacian}
\begin{align}
(D^2 \cA)_a :=
\big(\Box + \frac 12 \mu^2 - M^{(\cA)}_{rs}[\Theta^{rs},.]\big)^a_b   \cA_b \ .
\label{fluct-action}
\end{align}
Now consider the ansatz\footnote{The $x^\mu\frac{\k}{R}$ contribution is subleading here and dropped.} \eq{4D-modes-decomp}
\begin{align}
\cA_{\rm grav}^\mu(x) &= \frac{x^\mu}{R} \k + \theta^{\mu\nu} \big(A_{\nu\s}(x) P^\s + \tilde A_{\nu\s\r}(x)\cM^{\s\r}\big) , 
\qquad \cA^5 = \k
\end{align}
dropping the $U(1)$ component for now.
We can evaluate $D^2 \cA$ in the  semi-classical limit using the  basic rules 
\begin{align}
 [f(x),g(x)] &\sim i \theta^{\mu\nu} \del_\mu f \, \del_\nu g  \nn\\
 [\theta^{\mu\nu},g(x)] &\sim O(x \del g)|_p = 0   \nn\\
 [P_{\mu},g(x)] &\sim i\del_\mu g
 \label{semiclass-rules}
\end{align}
which are valid in the local ON frame at $p$. After some algebra (see appendix \ref{sec:eval-D2})
and 
\begin{align}
(\theta^{\mu\r}\theta^{\s\nu} -\theta^{\mu\s} \theta^{\r\nu}) \tilde A_{\nu\s\r} 
   = 2  r^2\theta^{\mu\nu} \cM^{\s\r} \tilde A_{\r\s\nu} 
\end{align}
using the antisymmetry of $A_{\nu\s\r}$  in the last two indices,
we obtain with  \eq{4D-modes-decomp}
\begin{align}
 D^2  \cA_{\rm grav}^\mu
  &= \theta^{\mu\nu} P^\s (\Box+8r^2 + \frac 12 \mu^2) A_{\nu\s}  +2 P^{\mu} \cM^{\s\nu} A_{\nu\s}  
  + 4 \g^{\mu\r} g^{\nu\s} \tilde A_{\nu\s\r}    \nn\\
  &\quad +\theta^{\mu\nu} \cM^{\s\r}\big(2r^2\del_\s A_{\nu\r} + (\Box +8r^2+ \frac 12 \mu^2) \tilde A_{\nu\s\r} 
   +  4 r^2 \tilde A_{\r\s\nu}  \big)  \nn\\
 D^2  \cA_{\rm grav}^5 &=  (\Box + 4r^2+\frac 12 \mu^2) \k \ .
 \label{D2-eval-full}
\end{align}
Now 
\begin{align}
 4\g^{\mu\r} g^{\nu\s} \tilde A_{\nu\s\r} 
    \ =  - 6r^2 \, \theta^{\mu\nu} \cM^{\s\r} P_0 \tilde A_{\nu\s\r}
\end{align}
where $A_{\nu\s\r}[\xi] = P_0 \tilde A_{\nu\s\r}$ is the trace contribution of $\tilde A_{\nu\s\r}$
defined in \eq{A-trace-xi}.
To evaluate $[\cA_{\rm grav} D^2 \cA_{\rm grav}]_{0}$, we need the following normalizations which follow from 
 \eq{PP-normaliz} and \eq{M-correlators-2sphere}
\begin{align}
  [P^\s  \theta^{\mu\nu}g_{\mu\mu'}\theta^{\mu'\nu'} P^{\s'} ]_{0} 
  &\ = L_R^2 \, g^{\nu\nu'} g^{\s\s'}  \nn\\
 [\cM^{\s\r} \theta^{\mu\nu}g_{\mu\mu'} \theta^{\mu'\nu'} \cM^{\s'\r'} ]_{0} 
   &= \ \frac{R^4}{3} \,  g^{\nu\nu'} (g^{\s\s'} g^{\r\r'} - g^{\s\r'} g^{\r\s'} +\epsilon^{\s\r\s'\r'})
\end{align}
using $\frac{\D^4}{4} = r^2 R^2$.
All other mixed terms such as $[P^\s  \theta^{\mu\nu}g_{\mu\mu'} \theta^{\mu'\nu'} \cM^{\s'\r'}]_{0}$ vanish.
Thus
\begin{align}
[P^\s A_{\nu\s} \theta^{\mu\nu} 
     \theta^{\mu'\nu'} A'_{\nu'\s'} P^{\s'} g_{\mu\mu'}]_{0} 
 &= L_R^2 \, A^{\nu\s} A'_{\nu\s} \nn\\
[\cM^{\s\r} A_{\nu\s\r} \theta^{\mu\nu} 
     \theta^{\mu'\nu'} A'_{\nu'\s'\r'} \cM^{\s'\r'} g_{\mu\mu'}]_{0} 
 &= \frac{4 R^4}{3} \, A^{\nu\s\r}  A'_{\nu\s\r} 
 \end{align}
 provided either $A_{\nu\s\r}$ or $A'_{\nu\s\r}$ is SD in the last 2 indices.
Therefore
the semi-classical  action \eq{fluct-action-nogf} quadratic in the $A_{\mu\nu}, \tilde A_{\nu\s\r}$ and $\k$ fields is
\begin{align}
S_2[\cA] &= \frac 1{g^2}\Tr \cA_a \Big((\Box + \frac 12 \mu^2)\d^a_b  + 2i [\Theta^{ab},. \, ] \Big) \cA_b  \nn\\
 &\sim \frac{4}{g_{\rm YM}^2 \D^8}
 \int_{\cM} \! \! d^4 x
  \Big(L_R^2\, A^{\mu\nu} (\Box + 8r^2+ \frac 12 \mu^2) A_{\mu\nu}
  + \k (\Box + 4r^2+ \frac 12 \mu^2) \k \nn\\
 &\qquad  + \frac{4 R^4}{3} \tilde A_{\nu\s\r}\big(\Box +8 r^2 + r^2 P_{\rm hor} - 6 r^2 P_0 + \frac 12 \mu^2\big) \tilde A^{\nu\s\r}
   + \frac{8 \a R^4}{3} \, r^2  \tilde A_{\nu\s\r} \del_\s A_{\nu\r} \ .
  \label{fluct-action-bare-explicit}
\end{align}
To accommodate one-loop effects (see section \ref{sec:one-loop}), we introduced a factor $\a$ in the mixed term, which is  
\begin{align}
 \a=1
\end{align}
for the present semi-classical action.
Given in addition a coupling to matter of the form\footnote{For scalar fields, the 
coupling of the trace component $h$ was found to be modified in \eq{scalar-coupling-linearized}.
This is somewhat puzzling; one possible resolution might be that the 
rescaling \eq{phi-varphi} should be done using the $x$-dependent uncertainty scale $\D^2_x$.
This should be addressed in more detail elsewhere.}
\begin{align}
 \d S[\phi] 
  &= \frac 12 \int_\cM d^4 x\, (h^{\mu\nu} + k^{\mu\nu}+\frac{2\k}{R}g^{\mu\nu}) T_{\mu\nu}
\end{align}
(cf. \eq{flux-coupling-linearized}, \eq{scalar-coupling-linearized})
we obtain the equations of motion
\begin{align}
  L_R^2 (\Box + 8r^2 + \frac 12 \mu^2) A_{\mu\nu} &= -\frac{g_{\rm YM}^2 \D^8}{16} T_{\mu\nu} 
          + \frac{\a R^2 \D^4}{3} \del^\s \tilde A_{\mu\s\nu}  \nn\\
 8R^2\big(\Box +8 r^2 + r^2 P_{\rm hor}- 6 r^2 P_0 + \frac{\mu^2}2\big)  A_{\nu\s\r}
   &= (P_{SD})^{\s'\r'}_{\s\r}\Big(- 2\a\D^4 \del_{\s'} A_{\nu\r'}
               + g_{\rm YM}^2\D^8 \del_{\s'} T_{\nu\r'}  \Big)  \nn\\
 2 (\Box + 4r^2+\frac 12 \mu^2) \k &= -\frac {g_{\rm YM}^2\D^8}{4R} T  \ .
 \label{eom-h-a-A3}
\end{align}
Here $P_{SD}^0 = \frac 1{4}(\d\d-\d\d+\epsilon)$ is the projector on the SD antisymmetric component\footnote{Strictly speaking this is 
no longer correct for the generalized spheres $\cS^4_\L$.
However the required generalizations would not significantly alter the conclusion, and we stick to the self-dual case here for simplicity.}, $P_0$ 
is the projector on the trace contributions of $\tilde A_{\mu\r\s}$, and
$P_{\rm hor}$ is the operator exchanging horizontal indices in the mixed hook diagram 
(corresponding to $\tilde A_{\mu\r\s}$)  defined in Appendix \ref{sec:young-projectors}. This
has eigenvalue
$P_{\rm hor} = -1$ on the totally antisymmetric diagrams, and $P_{\rm hor} = \frac 12$
on the mixed (hook) Young diagrams.
In any case, the contribution of $P_{\rm hor}$ and $P_0$ in the kinetic operator 
is negligible compared with $\Box$, and we neglect it for simplicity, and we replace the equation for $\tilde A_{\nu\s\r}$ by
\begin{align}
 R^2\big(\Box +8 r^2 + \frac{\mu^2}2\big) \tilde A_{\nu\s\r}
   &=  (P_{SD})^{\s'\r'}_{\s\r}\Big(- \frac{\a\D^4}4 \del_{\s'} A_{\nu\r'} + \frac {g_{\rm YM}^2\D^8}8 \del_{\s'} T_{\nu\r'}  \Big) \ .
\end{align}
This implies 
\begin{align}
 R^2\big(\Box +8 r^2 + \frac{\mu^2}2\big) \del^\s \tilde A_{\nu\s\r}
   &=  \frac{\a\D^4}{16} (-\del^\s\del_{\s} A_{\nu\r} + \del_\r\del^{\s} A_{\nu\s})
   + \frac{g_{\rm YM}^2\D^8}{32} \del^\s\del_{\s} T_{\nu\r} 
\end{align}
Now assume that the antisymmetric component of  $A_{\mu\nu}$ vanishes,  $a_{\mu\nu}=0$.
Then the gauge-fixing condition \eq{gaugefix-components} implies 
$\del^{\s} A_{\nu\s}= \frac 1R \del_\nu\k$, so that $\del^\s \tilde A_{\nu\s\r}$ is symmetric in $\nu,\r$.
Then the eom  \eq{eom-h-a-A3} for $A_{\mu\nu}$ is indeed consistent with 
\begin{align}
 a_{\mu\nu}=0, \qquad A_{\mu\nu} = \frac 12 h_{\mu\nu}
\end{align}
for symmetric and conserved $T_{\mu\nu}$.
Putting all this together and recalling $\Box \sim - \frac{\D^4}4\Box_g$,
we obtain the following equations for the metric contributions $k_{\mu\nu}$ \eq{kmunu-def-1}
and $h_{\mu\nu}$ and $\k$
\begin{align}
 \big(\Box_g - 4m^2\big) k_{\mu\nu}
   &= -\frac{g_{\rm YM}^2\D^4}3 \Box_g T_{\mu\nu} + \frac{\a}3 \Box_g h_{\mu\nu}
     - \frac{1}{3R} \del_\mu \del_\nu\k  \nn\\
  (\Box_g - 4m^2) h_{\mu\nu} &= \frac{g_{\rm YM}^2 \D^4}{2 L_R^2 } T_{\mu\nu} - \frac {\a}{L_R^2}  k_{\mu\nu} \nn\\ 
   (\Box_g - 4 m^2) \k &= \frac {g_{\rm YM}^2\D^4}{2R} T  \ .
  \label{eom-metrics}
\end{align}
Here we define the  mass 
\begin{align}
 m^2 = \frac{8r^2 + \mu^2/2}{\D^4} \ \stackrel{\mu^2 \to 0}{\approx} \  \frac{1}{4 R^2} \ 
\end{align}
which is just the  background curvature unless $\mu^2$ is significant.
We expect from \eq{eom-metrics} that  
$\frac{1}{R} \del_\mu \del_\nu\k = O(\frac {g_{\rm YM}^2\D^4}{R^2} T)  \ll g_{\rm YM}^2\D^4\Box_g T$,
so that we can neglect the $\k$ contribution (except possibly for the longest ``cosmological'' scales).
Then combining the first two equations gives
\begin{align}
 (\Box_g -4 m^2 + \frac {\a^2}{3L_R^2})  k_{\mu\nu}
   &= \frac{g_{\rm YM}^2\D^4}{3} (-\Box_g T_{\mu\nu} + \frac{\a}{2 L_R^2 } T_{\mu\nu}) 
     +\frac{4\a}{3} m^2\, h_{\mu\nu} \ .
     \label{equation-k}
\end{align}
It is now useful 
to separate  $k_{\mu\nu}$ 
into a ``local'' and a propagating ''gravitational`` part,
\begin{align}
 k_{\mu\nu} &= k_{\mu\nu}^{(loc)} + k_{\mu\nu}^{(grav)}  , \qquad
 k_{\mu\nu}^{(loc)} = -\frac{g_{\rm YM}^2\D^4}{3} T_{\mu\nu} \ .
  \label{k-loc-grav}
\end{align}
Then \eq{equation-k} reduces to 
\begin{align}
 (\Box_g  - m_k^2)  k_{\mu\nu}^{(grav)}  
   &= \frac{g_{\rm YM}^2\D^4}{3} \Big(\frac{\a}{2L_R^2} - m_k^2\Big) T_{\mu\nu}  
    +\frac{4\a}{3} m^2\, h_{\mu\nu} \ ,  \nn\\
   m_k^2 &:= 4m^2 -  \frac {\a^2}{3L_R^2}  \ .
     \label{equation-k-grav}
\end{align}
To ensure stability, we should require $m_k^2 \geq 0$, which means 
\begin{align}
 m^2 L_R^2 \geq  \frac {\a^2}{12} \ .
 \label{m2-LR-bound}
\end{align}
Moreover we want to assume that $m^2$ is very small but positive\footnote{As shown
in \cite{Steinacker:2015dra}, such a positive $m^2$ can in fact stabilize $S^4_N$. 
See section \ref{sec:one-loop} for more details.}. 
In view of \eq{LR-def}, this requires a generalized 4-sphere $\cS^4_\L$ with large $n$.
Then the term $m^2\, h_{\mu\nu}$ can be dropped, and
we can  solve \eq{equation-k-grav} as 
\begin{align}
 k_{\mu\nu}^{(grav)} 
  = \frac{\,g_{\rm YM}^2\D^4}{3L_R^2} \Big(\frac{\a}3 \big(\a + \frac{3}{2}\big) - 4L_R^2m^2\Big) \frac{1}{\Box_g  - m_k^2} T_{\mu\nu} \ .
\label{k-formal-solution}
 \end{align}
Inserting this in the equation for $h_{\mu\nu}$ gives 
\begin{align}
 h_{\mu\nu} 
  &= \frac{g_{\rm YM}^2 \D^4}{3 L_R^2 }\Big( \frac{\big(\a+\frac 32\big)}{\Box_g  - 4m^2} \, T_{\mu\nu}
   - \frac{ \frac {\a}{L_R^2} \Big(\frac{\a}3 \big(\a + \frac{3}{2}\big) - 4L_R^2m^2\Big) }{(\Box_g  - m_k^2)(\Box_g - 4 m^2)} T_{\mu\nu}\Big)  \ .
 \label{h-formal-solution}
\end{align}
As long as the masses $m^2$ and $m_k^2$ can be neglected,  
the first term is governed by the usual long-range $\frac 1{r^2}$ propagator (or rather Greens function), 
while the second term 
describes a sub-leading $\frac 1{r^4}$ correction\footnote{Observe that 
the second term is of order $\frac{m^2}{\Box_g^2}$ times the first term due to \eq{m2-LR-bound}.}.


Now we observe that the local contribution $k_{\mu\nu}^{(loc)} \sim T_{\mu\nu}$ \eq{k-loc-grav}
can be dropped; this  is merely a negligible local ''contact`` contribution to the metric,
and has nothing to do with any long-distance gravitational effect. 
For example, to compute the gravitational effect of the sun at the location of the earth, 
we certainly have $k_{\mu\nu}^{(loc)}=0$. 
Even for an observer located in a cloud of gas with some energy-momentum density $T_{\mu\nu}$,
the ''local`` contribution $k_{\mu\nu}^{(loc)} = O(g_{\rm YM}^2\D^4 \, T_{\mu\nu})$ 
would still be negligible except for extremely high energy densities.
Thus we will replace $\tilde h_{\mu\nu}$ by the  effective gravitational metric 
\begin{align}
 \tilde  h_{\mu\nu}^{(grav)} :=  h_{\mu\nu} + k_{\mu\nu}^{(grav)} \ .
 \label{tilde-h-grav}
\end{align}
Its physical significance  depends on the distance scale:

\paragraph{Regime G:}  Consider first the gravitational field for 
distance scales $D^2 \ll \frac 1{m^2}, \frac 1{m_k^2}$.
This is the ''gravity'' regime, where the propagators can be replaced by massless ones. 
Then the eom for $h_{\mu\nu}$ and $k_{\mu\nu}^{(grav)}$ reduce to 
\begin{align}
 \Box_g h_{\mu\nu} &= \big(\a+ \frac{3}{2}\big)\frac{g_{\rm YM}^2\D^4}{3 L_R^2}  T_{\mu\nu} \nn\\
 \Box_g  k_{\mu\nu}^{(grav)} &= \Big(\frac{\a}3 \big(\a + \frac{3}{2}\big) - 4L_R^2m^2\Big)\frac{g_{\rm YM}^2\D^4}{3 L_R^2}  T_{\mu\nu} \ ,
 \label{h-solution-G}
\end{align}
and the effective gravitational metric
satisfies 
\begin{align}
 \Box_g  \tilde h_{\mu\nu}^{(grav)} 
 &= \Big((\a +\frac 32)(\frac{\a}{3}+ 1\big) - 4L_R^2m^2 \Big)\frac{g_{\rm YM}^2\D^4}{L_R^2}  T_{\mu\nu} \ .
 \label{tilde-h-eq-G}
\end{align}
We will see that this amounts to the linearized Einstein equations.
Note that both equations are compatible with the 
Lorentz gauge condition $\del^\mu h_{\mu\nu} = 0 = \del^\mu k_{\mu\nu}$.
For $\a\to 0$, this regime always applies even for $L_R \approx 0$.

\paragraph{Regime C:} For distance scales $D^2 \gg \frac 1{m^2}, \frac 1{m_k^2}$, the mass in the
propagators becomes dominant.
Then both $k_{\mu\nu}^{(grav)}$ and $h_{\mu\nu}$ are exponentially damped, and 
$\tilde h_{\mu\nu}^{(grav)} \approx \tilde h_{\mu\nu}^{(loc)} = -\frac{g_{\rm YM}^2\D^4}{3} T_{\mu\nu}$.
This is the ``cosmological'' regime.
Hence gravity ceases to operate for structures much larger than $m^2$, as always in massive theories.

\paragraph{Intermediate regime.}
Finally, there may be an interesting intermediate regime provided $m^2 \gg m_k^2$, 
for distances  $\frac 1{m^2} \ll D^2 \ll  \frac 1{m_k^2}$.
Then  $h_{\mu\nu}$ is exponentially damped, while $k_{\mu\nu}^{(grav)}$ is still in its massless, propagating regime.
Then we still get the Einstein equations
\begin{align}
  \Box_g  \tilde h_{\mu\nu}^{(grav)} 
 &= \Big(\frac{\a}{3}(\a +\frac 32) - 4L_R^2m^2 \Big)\frac{g_{\rm YM}^2\D^4}{L_R^2}  T_{\mu\nu} \ ,
 \label{tilde-h-eq-G-intermediate}
\end{align}
but the effective Newton constant is reduced compared with \eq{tilde-h-eq-G}.
Formally, \eq{equation-k-grav} admits the possibility that  $m^2 \gg m_k^2 \approx 0$ even for small $L_R^2$. 
Then this intermediate regime would provide long-range gravity even for an almost-basic 
$\cS^4_\L$ with $n \approx 0$. However this would be fine-tuning, and it seems unlikely that this is
compatible with the radial stabilization, cf. \cite{Steinacker:2015dra} and section \ref{sec:one-loop}.

\paragraph{Self-dual action.}
We conclude this section with the following interesting observation.
Suppose the matrix model has an additional $\epsilon$ term such that it reduces to 
a selfdual Yang-Mills action\footnote{It was found in \cite{Steinacker:2010rh,Steinacker:2007dq} 
that the self-dual action indeed arises upon taking into account the 
volume factor contributed by the fluctuations, cf. \eq{cA-expand-e}. We leave this  possibility 
for future work.} 
\begin{align}
 \Gamma_{SD}[\cF^2]  =
  \int\limits d^4x\, \big[\cF^{\mu\nu}_+(\xi) \cF_{\mu\nu}^+(\xi)\big]_{0} 
  \label{action-SD}
\end{align}
where 
\begin{align}
 \big[\cF_+^{\mu\nu} \cF^+_{\mu\nu}\big]_{0} 
 &= \big[2\cF^{\mu\nu} \cF_{\mu\nu}  + \cF^{\mu\nu} \cF^{\r\s} \epsilon_{\mu\nu\r\s}\big]_{0} \ 
\end{align}
is averaged over the local fiber, and
\begin{align}
 \cF^{\mu\nu} = 
 \theta^{\mu \mu'}   A_{\mu'\nu'} g^{ \nu'\nu} - \theta^{\nu \nu'}  A_{\nu'\mu'}g^{\mu' \mu} 
  + \theta^{\mu\mu'} \theta^{\nu \nu'} F_{\mu'\nu'}[\A] 
\end{align}
cf. \eq{flux-perturbed}.
The first term is equivalent (modulo gauge fixing) to the quadratic action \eq{fluct-action-bare-explicit}. 
The second term $\cF^{\mu\nu} \cF^{\r\s} \epsilon_{\mu\nu\r\s}$ as usual 
will not affect the local 
field equations, except for the mixed term between $A_{\mu\nu}$ and $F_{\mu\nu}$, which after some algebra is
\begin{align}
 &  F_{\mu'\nu'}[\A] \theta^{\mu\mu'} \theta^{\nu \nu'}\epsilon_{\mu\nu\r\s}
(\theta^{\r \r'}   A_{\r'\s'} g^{ \s'\s} - \theta^{\s \s'}  A_{\s'\r'}g^{\r' \r} )  
 = -\frac{16 R^4 r^2}{3} \, \del^\mu \tilde A_{\a\mu\r}  A^{\r\a} 
   \label{FFeps-action}
\end{align}
 using selfduality of $A_{\nu\a\b}$, its symmetry property and
 $\det\theta 
 = (\frac{\D^4}{4})^2$.
This has indeed the same form as the mixed term in \eq{fluct-action-bare-explicit} (which
is the reason for introducing the factor $\a$ in \eq{fluct-action-bare-explicit}). 
Similarly, 
\begin{align}
  & F_{\mu'\nu'}[\A] \theta^{\mu\mu'} \theta^{\nu \nu'}g_{\mu\r}g_{\nu\s}
(\theta^{\r \r'}   A_{\r'\s'} g^{ \s'\s} - \theta^{\s \s'}  A_{\s'\r'}g^{\r' \r} )   
 \  \sim \  \frac{8 R^4 r^2}{3} \del^\mu \tilde A_{\r\mu\a} (A^{\r\a} + \frac{\k}{R} g^{\r\a} )     \nn  
\end{align}
using partial integration and the gauge fixing condition. The contribution from $h_{\r\a}$ 
agrees up to a factor (-2) with \eq{FFeps-action}. Thus the mixed term cancels in the
selfdual Yang-Mills action, which
reduces to \eq{fluct-action-bare-explicit} with $\a=0$.  Then
the regime G always applies, without the need for large $L_R$. Together with the following section this implies 
the linearized Einstein equations always arise, without IR modification.
We leave this as an interesting observation.

\subsection{Curvature and linearized Einstein equations}

Now we consider the curvature of the linearized effective metric 
\begin{align}
 G^{\mu\nu} = g^{\mu\nu} + H^{\mu\nu} = \d^{\mu\nu} + \d g^{\mu\nu} + H^{\mu\nu}
\end{align}
viewed as a perturbation of the flat metric $\d^{\mu\nu}|_p = g^{\mu\nu}|_p$ near $p$;
recall that $H^{\mu\nu}$ was defined in \eq{H-def} as trace-reverse of 
$\tilde h_{\mu\nu} = h_{\mu\nu} + k_{\mu\nu}$, or rather by $\tilde h_{\mu\nu}^{(grav)}$ 
\eq{tilde-h-grav} as discussed above.
Furthermore, assume that we are in the scaling regime G, where the masses $m^2$ and $m^2_k$ can be neglected.
Then consider the linearized Ricci tensor \cite{Wald:1984rg}
\begin{align}
 R^{\mu\nu}[g+H] \ &\approx \    R^{\mu\nu}[g] +  \frac 12 \del^\mu\del^\nu H + \frac 12 \del^\a\del_\a H^{\mu \nu} 
 - \del^{(\mu}\del_\r H^{\nu)\r} \ .
\end{align}
Since $H_{\mu\nu}$ satisfies the de Donder gauge \eq{dedonder},
it simplifies as
\begin{align}
 R_{\mu\nu}[g+H] \ &\approx \ \frac{3}{R^2} g_{\mu\nu} +\frac 12 \del^\a\del_\a H_{\mu \nu} , 
\end{align}
where 
$R_{\mu\nu}[g] = \frac{3}{R^2} g_{\mu\nu}$ is the  Ricci tensor of $S^4$. Hence
the linearized Einstein tensor is
\begin{align}
 \cG_{\mu\nu}[g+H]  \ &=  \  R_{\mu\nu}[g+H]   - \frac 12 g_{\mu\nu}  R[g+H] 
  \ \approx \  \frac 12 \Box_g \tilde h_{\mu\nu}^{(grav)}
\end{align}
dropping the curvature contribution  $\cG[g] = -\frac{3}{R^2} g_{\mu\nu}$  of $S^4$.
Taking into account the equation of motion 
\eq{tilde-h-eq-G} for $\tilde h_{\mu\nu}^{(grav)}$,
 we obtain the linearized Einstein equations
\begin{align}
\boxed{
  \ \cG_{\mu\nu}   \ = \  8\pi G_N T_{\mu\nu}   \ 
   }
 \label{Einstein}
\end{align}
with the Newton constant given by 
\begin{align}
  G_N =  \Big((\a +\frac 32)(\frac{\a}{3}+ 1\big) - 4L_R^2m^2 \Big)\frac{g_{\rm YM}^2\D^4}{48\pi L_R^2} \  =: L_{pl}^2 \ 
  \ \leq  \ O(g_{\rm YM}^2 \D^2)
 \label{Planck-scale}
\end{align}
using \eq{LR-def}. This entails the interesting reciprocity relation\footnote{This admits
the interesting possibility that $L_{pl}^2 \ll \D^2$, which would mean that the scale of noncommutativity $\D^2$ can be much larger 
than the Planck scale.}
\begin{align}
 L_{pl}^2 L_R^2 = O(g_{\rm YM}^2 \D^4) \ .
\end{align}
Since the gravity regime G applies  only for distances smaller than $\frac 1m \leq  L_R/\a$ \eq{m2-LR-bound},
we should  require $L_{pl} \ll L_R/\a$. Therefore one of the following conditions must hold
\begin{align}
c_n \gg  \sqrt{N}  \qquad \mbox{or} \qquad \a\approx 0 \ .
\label{cn-estimate-gravity}
\end{align}
Thus we need either  a self-dual action, or a very ''thick`` fuzzy sphere $\cS^4_\L$.
Notice that such a macroscopic ``thickness'' $L_R$ of $\cS^4_\L$ does not necessarily mean that space 
 is effectively 5-dimensional. This point was discussed in section \ref{sec:dimred},
although a more detailed investigation  is required to settle this. 
There is no such issue with dimensional reduction for the self-dual action \eq{action-SD}, 
where $c_n$  can be very small.

%
%

 Now consider briefly 
the gravitational coupling of fermions.
We only observe here that the matrix model  Dirac operator 
\begin{align}
 \slashed{D}\Psi =  \Gamma_a [X^a,\Psi] \ \sim \ i(\tilde \g^\mu \del_\mu + ...)\Psi
\end{align}
can be 
rewritten in terms of ``comoving'' Clifford generators 
$\tilde \g^\mu = \Gamma_\a\, \tilde e^{\a\mu}$ (up to possibly a conformal factor),
which encode  the effective metric $\{\tilde \g^\mu,\tilde \g^\nu\} = 2 G^{\mu\nu}$
 \cite{Steinacker:2010rh}. 
Together with supersymmetry \cite{Ishibashi:1996xs}, we expect that  fermions  couple 
properly to gravity in the present framework, however
a detailed analysis is left for future work.

The above results show that  the 4-dimensional 
(Euclidean, linearized) Einstein equations can emerge from 
the classical dynamics of fluctuations on  fuzzy 4-spheres $\cS^4_\L$  in the 
Yang-Mills matrix model \eq{bosonic-action}, provided certain  conditions 
for $\L$ are met (i.e. large $n$), and dimensional reduction is justified.
No explicit Einstein-Hilbert term is required. 
There are several contributions to the metric fluctuation $\tilde h_{\mu\nu}$, so 
that the physics is  richer than in general relativity. Most notably there is a long-distance cutoff of gravity
set by some IR mass scale $m^{-1}$.
The requirement of large $n$ might be avoided for 
the self-dual action.

In any case, the long-wavelength modifications of gravity discussed above 
could be very interesting for cosmology, and 
it is tempting to relate this to some of the effects attributed to dark matter or dark energy. 
There will also be some additional gravitational modes arising e.g. from radial deformations $\k$. 
However, a more detailed analysis is required before these issues can be addressed.

One final but crucial question is whether the present mechanism survives quantization.
The (preliminary and partial) analysis in the following section supports the conjecture that 
the quantization is well-behaved and preserves the above picture, for the IKKT model.

\section{One-loop corrections from string states}
\label{sec:one-loop}

As a first step towards a full quantum theory,
we would like to study the one-loop effective action for the above gravitational fluctuations around a 
fuzzy $\cS^4_\L$ background. This should  be done in the maximally supersymmetric IIB or IKKT model,
which is the only model where the non-local UV/IR mixing in noncommutative field theory is mild
(leading to 10-dimensional IIB supergravity in target space). 
Until recently, such a 1-loop computation in terms of a  mode expansion 
would be hopeless; already the one-loop effective action for the constant radius  is quite 
involved \cite{Steinacker:2015dra}.
However the integration method using string states  \cite{Steinacker:2016xx} makes this task feasible.
As a check of these methods 
we will first reproduce the results in \cite{Steinacker:2015dra}, and then 
proceed to extract the leading 1-loop contributions to the effective action.
For simplicity we will restrict ourselves to the basic fuzzy sphere $\cS^4_N$ here.

\subsection{The 1-loop effective potential for the IKKT model}
\label{sec:one-loop-pot}

We start with the bare bosonic action \eq{bosonic-action} for the background $X$ 
\begin{align}
  S_0[X] &= \frac 1{g^2}\Tr \Big(-[X_i,X_j][X^i,X^j]\, + \mu^2 X^i X_i \Big) 
\end{align}
supplemented by a mass $\mu^2$, to regularize possible IR singularities. 
Adding the fermions in the IKKT model,
the one-loop  effective action is defined by
\begin{align}
 Z[r,\mu] &= \int\limits_{\rm 1\, loop} dX d\Psi e^{-S[r \bar X,\Psi]} 
 = e^{- \Gamma_{\!\textrm{eff}}[r,\mu] } \
 \end{align}
and we will write 
 \begin{align}
 \Gamma_{\!\textrm{eff}}[r,\mu] &= S_0[X] + \Gamma_{\!\textrm{1loop}}[r,\mu] \ .
\end{align}
We recall  
the following form of the one-loop effective action in the 
IKKT model \cite{Ishibashi:1996xs,Chepelev:1997av,Blaschke:2011qu} 
\begin{align}
\Gamma_{\!\textrm{1loop}}[X]\! &= \frac 12 \Tr \Big(\log(\Box +\frac{\mu^2}2 - M^{(\cA)}_{ab}[ \Theta^{ab},.])
-\frac 12 \log(\Box - M^{(\psi)}_{ab}[ \Theta^{ab},.])
- 2 \log (\Box)\Big)   \nn\\
 &= \frac 12 \Tr \Bigg(\sum_{n>0} \frac{1}n \Big((\Box^{-1}\big(-M^{(\cA)}_{ab}[ \Theta^{ab},.] 
    + \frac{1}{2}\mu^2)\big)^n 
  \, -\frac 12 (-\Box^{-1}M^{(\psi)}_{ab}[ \Theta^{ab},.])^n \Big)  \Bigg) \nn\\
  &= \frac 12 \Tr \Bigg(\!\! \frac 14 \Box^{-1}(M^{(\cA)}_{ab} [ \Theta^{ab},.] )^4 
  -\frac 18 (\Box^{-1}M^{(\psi)}_{ab} [ \Theta^{ab},.])^4 \,\, +  \cO(\Box^{-1}[ \Theta^{ab},.])^5 \! \Bigg) \nn\\
  &\quad + \frac 14  \mu^2 \Tr \Box^{-1} + O(\mu^4)
\label{Gamma-IKKT}
\end{align}
with  $a,b=1,...,10$,
where
\begin{align}
\begin{array}{rl}(M_{ab}^{(\psi)})^\a_\b &= \frac 1{4i} [\Gamma_a,\Gamma_b]^\a_\b \,  \\
                      (M_{ab}^{(\cA)})^c_d &= i(\d^c_b \d_{ad} - \d^c_a \d_{bd}) \, , \\
                    \end{array}  
\end{align}
and the  $2 \log \Box$ term arises from the ghost contribution.
Note that the coupling constant $g$ drops out  from $\Gamma_{\!\textrm{1loop}}$
due to supersymmetry.
For $\mu=0$, the first non-vanishing term in this expansion is  $n=4$ due to maximal supersymmetry.
However there are contributions of order $\Theta$ for $\mu^2\neq 0$ due to the soft SUSY breaking, 
which are important to stabilize the background.

The leading 4th order term
is given by the following expression \cite{Blaschke:2011qu}:
\begin{align}
 \Gamma_{\!\textrm{1loop};4}[X]\! &=
 \frac 18 \Tr \Bigg(\!\!  (\Box^{-1}(M^{(\cA)}_{ab} [ \Theta^{ab},.])^4 
  -\frac 12 (\Box^{-1}M^{(\psi)}_{ab} [ \Theta^{ab},.])^4 \Bigg) \nn\\
  &= \frac 14 \Tr\Big(\Box^{-1}[\Theta^{a_1 b_1}, \ldots \Box^{-1}[\Theta^{a_4 b_4},.]]]]\Big) \nn\\
 &\quad  \big(-4 g_{b_1 a_2} g_{b_2 a_3} g_{b_3 a_4} g_{b_4 a_1} 
- 4 g_{b_1 a_2} g_{b_2 a_4} g_{b_4 a_3} g_{b_3 a_1} 
- 4 g_{b_1 a_3} g_{b_3 a_2} g_{b_2 a_4} g_{b_4 a_1} \nn\\
&\quad +  g_{b_1 a_2} g_{b_2 a_1} g_{b_3 a_4} g_{b_4 a_3}
 +  g_{b_1 a_3} g_{b_3 a_1} g_{b_2 a_4} g_{b_4 a_2}
+  g_{b_1 a_4} g_{b_4 a_1} g_{b_2 a_3} g_{b_3 a_2} \big) 
\end{align}
and the leading term in $ \mu^2$ is
\begin{align}
 \Gamma_{\!\textrm{1loop}}[X;\mu^2]\! &= - \frac 14 \mu^2 \Tr\big( \Box^{-1}\big) \ .
\end{align}
We will evaluate the trace  over hermitian matrices in $End(\cH)$
 using the basic formula \cite{Steinacker:2016xx}
\begin{align}
 \Tr_{End(\cH)} \cO &= \frac{(\dim \cH)^2}{(\rm Vol \cM)^2}\,\int\limits_{\cM\times\cM} d{\bf x} d{\bf y}
  (|{\bf x}\rangle\langle {\bf y}|) \cO (|{\bf y}\rangle\langle {\bf x}|) \ .
  \label{trace-coherent-End-hermit}
\end{align}
Here $|{\bf y}\rangle\langle {\bf x}|\in End(\cH)$ are string states\footnote{Bold face letters 
${\bf x}, {\bf y}, ...$
denote points in $\C P^3$, while plain letters $x,y,...$ denote their projection on $S^4$.}, 
which are built out of coherent states $|{\bf x}\rangle$
on $\cM = \C P^3 \approx \cS^4_N \times S^2$.
The formula is exact for homogeneous spaces such as $\C P^3$. It follows by noting that 
the rhs of \eq{trace-coherent-End-hermit} is invariant under $SO(5)_L\times SO(5)_R$, and so is 
the trace functional on $End(\cH_\L)$.
The normalization of the measure\footnote{The proper measure is the symplectic volume form on the underlying $\C P^3$.} 
in the integrals cancels out, and we will choose it as product measure on $S^2\times S^4$ 
with unit volume of $S^2$  and the  measure on $S^4$ is induced by the target space metric.
For deformed $\cM$, \eq{trace-coherent-End-hermit} is expected to be an excellent approximation, 
as long as  $\cO$ is well localized.

The string states are very useful for loop computations, because they
have approximate localization properties in {\em both} position and momentum; 
see \cite{Steinacker:2016xx} for a detailed discussion.
In particular, 
\begin{align}
 \Box^{-1}(| {\bf y}\rangle\langle {\bf x}|) &\sim \frac 1{|x- y|^2 + 2\Delta^2} \ | {\bf y}\rangle\langle {\bf x}|  \nn\\
 \Box^{-1}[ \Theta^{ab},.] (| {\bf y}\rangle\langle {\bf x}|) 
 &\sim \frac 1{|x- y|^2+ 2\Delta^2}\, \d\Theta^{ab}( {\bf y},{\bf x}) | {\bf y}\rangle\langle {\bf x}| \nn\\
 \d\Theta^{ab}( {\bf y},{\bf x}) &=  \Theta^{ab}( {\bf y})-  \Theta^{ab}({\bf x}) \ .
\end{align}
We can therefore approximately evaluate the 1-loop integral as follows
\begin{align}
\Gamma_{\!\textrm{1loop};4}[X]\! 
 &=  \frac 14 \frac{(\dim\cH)^2}{(\rm Vol(\cM))^2}\,
  \int\limits_{\cM\times\cM} d{\bf x} d {\bf y}  
  \frac{ \d\Theta^{a_1b_1}( {\bf y},{\bf x})  
  \d\Theta^{a_2b_2}( {\bf y},{\bf x}) \d\Theta^{a_3b_3}( {\bf y},{\bf x})  \d\Theta^{a_4b_4}( {\bf y},{\bf x})}
  {(|{x}- y|^2+2\Delta^2)^4}  \nn\\
 &\quad 3\big(-4 g_{b_1 a_2} g_{b_2 a_3} g_{b_3 a_4} g_{b_4 a_1} 
  + g_{b_1 a_2} g_{b_2 a_1} g_{b_3 a_4} g_{b_4 a_3} \Big)  \nn\\
  &= \frac 14 \frac{(\dim\cH)^2}{(\rm Vol(\cM))^2}\,
  \int\limits_{\cM\times\cM} d{\bf x} d {\bf y} 
   \frac{3 S_4[ \d\Theta({\bf x},{\bf y})] }{(|x-y|^2+2\Delta^2)^4}  \nn\\
 \Gamma_{\!\textrm{1loop}}[X;\mu^2]\! &= 
   \frac{5}2 \frac{(\dim\cH)^2}{(\rm Vol(\cM))^2}\,\int\limits_{\cM\times\cM} d{\bf x} d {\bf y} 
   \frac{\mu^2}{|x-y|^2+2\Delta^2} \ 
\label{1-loop-coh}
\end{align}
where
\begin{align}
 S_4[ \d\Theta] = -4tr \d\Theta^4 + (tr \d\Theta^2)^2 \ 
\end{align}
(suppressing the target space metric $g_{ab}$).
First, we note that $\Gamma_{\!\textrm{1loop};4}[X]$ vanishes identically for 
constant fluxes $\Theta = const$. This is a reflection of the maximal supersymmetry
of such backgrounds. Due to the SUSY cancellations, the interaction 
decays like $r^{-8}$, and it is bounded 
at short distances by the NC cutoff $\Delta^2$. This means that the one-loop induced action 
is a  weak short-distance effect on branes with dimension less than 10 
(which is essentially IIB supergravity).
We will compute its  effect on the fluctuations on the fuzzy $S^4$ background below.

The following observation \cite{Tseytlin:1999dj} is very useful:
If  $ \d\Theta^{ab}({\bf y},{\bf x})$ has rank $\leq 4$ 
for any fixed points ${\bf y},{\bf x}$ (which holds for any geometries embedded in $\R^5$), then  
\begin{align}
 - S_4[ \d\Theta] &= 
   4  tr(\d\Theta g  \d\Theta g  \d\Theta g  \d\Theta g) - (tr \d\Theta g  \d\Theta g)^2 \nn\\
  &=  4 (\d\Theta_+^{ab}{ \d\Theta_+}_{ba})\, ( \d\Theta_-^{cd}{ \d\Theta_-}_{dc}), \qquad 
  \d\Theta_\pm =  \d\Theta \pm \star_g  \d\Theta \, \nn\\
   &\geq 0
\label{F+F-indentity}
\end{align}
where $\star_g$ denotes the 4-dimensional Hodge star with respect to  $g_{\mu\nu}$.
Hence  $S_4$  leads to an {\em attractive} interaction, which vanishes
only in the (anti-) selfdual case $ \d\Theta = \pm \star_g \d\Theta$. 
This means that the quantum effects on fuzzy $S^4$ are small, because $\theta^{\mu\nu}$ is self-dual here.

\subsection{Vacuum energy of $\cS^4_N$}

\paragraph{Mass contribution.}

We start with the contribution of $\mu^2$ \eq{1-loop-coh}:
\begin{align}
  \Gamma_{\!\textrm{1loop}}[X;\mu^2]\!
  &= \frac 52 \frac{(\dim\cH)^2}{(\rm Vol(\cM))^2}\,\int\limits_{\cM\times\cM} d{\bf x} d{\bf y}
  \, \frac{\mu^2}{|x-y|^2+2\D^2}  \nn\\
  &= \frac 52  \frac{(\dim\cH)^2}{\rm Vol(S^4)}\,\int\limits_{S^4} d x\,
   \frac{\mu^2}{R^2|x-p|^2+2\D^2}   \nn\\
 &= \mu^2 \frac 52  \frac{(\dim\cH)^2}{R^2 \frac{8\pi^2}{3}}\,\int\limits_{0}^{\pi} 2\pi^2d\vartheta 
   \frac{\sin^3\vartheta}{(1-\cos\vartheta)^2 + \sin\vartheta^2 + 2\tilde\Delta^2}   \nn\\ 
 &=  \frac{\mu^2}{r^2}\frac {15}{8}\frac{(\dim\cH)^2}{R_N^2}\,
  \big(1 + O(\tilde\D^2) \big) 
\end{align}
where
 \begin{align}
 \tilde\Delta^2 \sim \frac{\D^2}{R^2} = \frac{4}{N} 
 \label{tilde-delta}
\end{align}
using \eq{Delta-S4}.
Note that $S^4$ denotes the unit sphere in this computation.  
Using 
\begin{align}
 \dim\cH = \frac 16 (N + 1)(N + 2)(N + 3) ,
\end{align}
we obtain 
\begin{align}
  \Gamma_{\!\textrm{1loop}}[X;\mu^2]\!
   &= \frac {5}{24} N^4\, \frac{\mu^2}{r^2} \big(1 + O(\frac 1{N}) \big) \ .
\label{gamma-mu-final}
\end{align}
This agrees precisely with the group-theoretical computation in 
\cite{Steinacker:2015dra}. This term describes the positive vacuum energy contribution due to the explicit 
SUSY breaking by the bosonic mass  $\mu^2$, which scales like $\frac{\mu^2}{r^2}$.

\paragraph{Curvature contribution. }

Now we compute the $\mu^2=0$ contribution  
\begin{align}
\Gamma_{\!\textrm{1loop};4}[X]\! 
  &= \frac 14 \frac{(\dim\cH)^2}{(\rm Vol(\cM))^2}\,
  \int\limits_{\cM\times\cM} d{\bf x} d{\bf y}
   \frac{3 S_4[ \d\Theta({\bf x},{\bf y})] }{(|x-y|^2+2\D^2)^4} \ .
   \label{gamma-S4-2}
\end{align}
We can fix ${\bf x} = (x,\xi)$ to be some fixed reference point on $\cM \approx S^4\times S^2$,
where $x^\mu$ are local coordinates on $S^4$.
We first compute the integration over $\eta \in S^2$  with $y=x$, 
which is the projection defined in \eq{average-def}
\begin{align}
 \left[ f(x,\eta) \right]_{S^2} := \frac{1}{4\pi}\int\limits_{S^2} d\eta f(x,\eta) 
 \end{align}
at any given  $x\in S^4$. 
Recalling the identity $\g^{bc} = \frac 14 \D^4  P^{bc}_T(x)$ 
 where $P_T$ is the tangential projector on $S^4 \subset \R^5$ and using
 \eq{M-correlators-2sphere}, this gives
\begin{align}
 \left[ \theta^{ab}_{\bf x} \theta^{cd}_{\bf x}\right]_{S^2} 
  &=\frac{1}{12} \D^4   (P^{ac}_T(x) P^{bd}_T(x) - P^{bc}_T(x) P^{ad}_T(x) + \varepsilon^{abcde} x^e) \nn\\
 \left[ \g^{ab}_{\bf x} \g^{bc}_{\bf y}\right]_{S^2\times S^2} 
  &= \frac 1{16} \D^8  (g^{ac} - y^a y^c - x^a x^c + x^a (x\cdot y) y^c)  \nn\\
  \left[ tr \g_{\bf x} \g_{\bf y}\right]_{S^2\times S^2} 
  &= \frac 1{16} \D^8 P^{ab}_T(x) P^{ba}_T(y)   
  =  \frac 1{16} \D^8  (3 + (x\cdot y)^2) \nn\\
  \left[ tr(\theta_{\bf x}\theta _{\bf y}\theta_{\bf x}\theta_{\bf y})\right]_{S^2\times S^2} 
        &= \frac 1{144} \D^8  \big((3+(x\cdot y)^2)(2+(x\cdot y)^2) -24 (x\cdot y) \big) 
       \  \stackrel{x\to y}{\to} \  -\frac{1}{12} \D^8  \nn\\
  \left[(tr(\theta_{\bf x} \theta_{\bf y}))^2 \right]_{S^2\times S^2} 
        &= \frac 1{144} \D^8 \big(2(3+(x\cdot y)^2)(2+(x\cdot y)^2) +24 (x\cdot y) \big) 
       \  \stackrel{x\to y}{\to} \ \frac{1}{3} \D^8 \   
 \label{M-correlators-2sphere-5D}
\end{align}
using the notation $\theta^{ab}_{\bf x} = \theta^{ab}(x,\eta)$ etc.,
and $x\cdot y= x^b y_b$,   and 
\begin{align}
 P^{ab}_T(x) P^{ab}_T(y) &= 3+(x\cdot y)^2  \nn\\
  P^{ab}_T(x) P^{bc}_T(y) P^{cd}_T(x) P^{da}_T(y) 
  &= 3 + (x\cdot y)^2 \nn\\
 \varepsilon^{abcde} x_e \varepsilon_{abcdf} y^f &= 24 (x\cdot y) \ .
\end{align}
We can now evaluate \eq{gamma-S4-2}  as follows
\begin{align}
\Gamma_{\!\textrm{1loop};4}[X]\! 
  &= \frac 34 \frac{(\dim\cH)^2}{(\rm Vol(\cM))^2}\,
  \int\limits_{\cM\times\cM} d {\bf x} d{\bf y}\frac{1}{(|x-y|^2+2\D^2)^4}  
  \Big[ -4tr(\g_{\bf x}^2)  + (tr \g_{\bf x})^2 + ({\bf x}\leftrightarrow {\bf y})  \nn\\
 &\quad  +4\big(4tr(\g_{\bf x} \theta_{\bf x} \theta_{\bf y}) 
 - tr(\theta_{\bf x} \theta_{\bf y})tr (\g_{\bf x}) + ({\bf x}\leftrightarrow {\bf y}) \big) \nn\\
 &\quad - 16tr(\g_{\bf x} \g _{\bf y})  + 2 tr \g_{\bf x} tr \g_{\bf y} 
 - 8 tr(\theta_{\bf x}\theta_{\bf y} \theta_{\bf x}\theta_{\bf y}) 
  + 4 (tr(\theta_{\bf x} \theta_{\bf y}))^2 \Big]_{S^2\times S^2}  \nn\\
 &= - \frac 3{4}  \frac{(\dim\cH)^2}{(\rm Vol(S^4))^2}\,\frac{\D^8}{R^8} 
  \int\limits_{S^4\times S^4} dx d y \frac{\big(1 - (x\cdot y)\big)^2}{(|x-y|^2+2\tilde\D^2)^4}  \nn\\ 
 &= - \frac 1{4} N^2\, \Big(-\frac{17}{3} + 2\ln 2 - 2 \ln \tilde \D^2 + \cO(\tilde\D^2) \Big) \nn\\ 
 &= - \frac 1{2} N^2\, \Big( \ln N + O(1) + \cO(\tilde\D^2) \Big) \ .
 \label{gamma-S4-final}
\end{align}
Again the last line agrees  with the (more involved) group-theoretical computation
in \cite{Steinacker:2015dra}, providing further support for the 
coherent state approach.
The present computation is not only shorter, it also allows
to see more clearly the origin of the attractive interaction:
It arises from nearly-local interaction among the $N$ degenerate sheets at  points 
$x \approx y \in S^4$.
At coincident locations $x=y$, the cancellation is exact,  because  
$\theta^{\mu\nu}(x,\xi)$ is selfdual.
In other words, the interaction is a residual attractive IIB supergravity effect which arises
due to the curvature of $S^4$.
This also confirms the stabilization mechanism put forward in \cite{Steinacker:2015dra}.

\subsection{Fluctuations}

Having gained confidence in the coherent state approach to  1-loop integrals on fuzzy $S^4$, 
we turn to the 1-loop effective action for the fluctuation fields.
This is of course a major task, and  we will only consider the leading  corrections to the kinetic terms for the 
lowest spin excitations of interest here.

\subsection{Transversal fluctuations}

Consider first the contributions from the transversal flux components 
$\cF^{\mu a} \sim -i[X^\mu,\cA^a]$ where $a=5,...,10$.
It is easy to see that only the transversal fluctuations $\phi^i$ in 
\begin{align}
 X^a = \begin{pmatrix}
        x^\mu \\ \phi^i
       \end{pmatrix}
\end{align}
 contribute to $\cF^{\mu a}$.
The general formula \eq{F+F-indentity} for $S_4[\d\theta+\d\cF]$ shows that the 
dominant interactions arise for points ${\bf x} = (x,\xi)$ and ${\bf y} = (x,\eta)$ on $\cM= \C P^3$ 
with the same $x\in S^4$. 
Then transversal fluctuations can arise only at quadratic (or higher) order,  contracted as
\begin{align}
 \d\cF^{\mu a} g_{ab}\d\cF^{b\nu} &= (\cF^{\mu a}(x,\xi)-\cF^{\mu a}(x,\eta)) g_{ab} (\cF^{b\nu}(x,\xi)-\cF^{b\nu}(x,\eta)) \nn\\
   &= \d\theta^{\mu \a} \,\tilde T[\phi]_{\a\b} \,\d\theta^{\b\nu} \ .
 \end{align} 
 Here we assume that $\phi^a = \phi^a(x)$ is constant along the $S^2$ fiber,
 so that $\cF^{\mu a} = \theta^{\mu\a} \del_\a \phi^a$ \eq{D-phi-S4}, and
 \begin{align}
 \tilde T[\phi]_{\a\b} &=  \del_\a \phi^a \del_\b \phi_a, \qquad
 \quad \d\theta^{\mu\a} = \theta^{\mu \a}(x,\xi) - \theta^{\mu \a}(x,\eta) \ .
\end{align}
However, we claim that this quadratic contribution in $\phi$ cancels due to the averaging over $S^2$,
and the only non-vanishing contributions are higher-order interactions or  higher-derivative terms. 
To see this, note that the quadratic contribution would arise from 
\begin{align}
 S_4[\phi] &= 
  - 4 tr((\d\theta +\d\cF)^4) + (tr (\d\theta  + \d\cF)^2)^2 \nn\\
  &= -16   tr( \d\theta\d\theta\d\theta\d\theta \tilde T[\phi]) 
  + 4 \, tr(\d\theta\d\theta) tr(\d\theta\d\theta \tilde T[\phi]) \nn\\
  &\quad + O(\phi^4) + O((\del^2\phi)^2)  \ .
\label{F+F-indentity-transversal}
\end{align}
Averaging over $(\xi,\eta) \in S^2\times S^2$  and using
invariance under $SU(2)_L$ and therefore under $SO(4)$ (noting that $\tilde T[\phi]$ is constant on $S^2$) gives
\begin{align}
 \left[(\d\theta\d\theta\d\theta\d\theta)^{\mu\nu} \right]_{S^2\times S^2}
  &= \frac 14 g^{\mu\nu} [tr(\d\theta\d\theta\d\theta\d\theta)]_{S^2\times S^2}  \nn\\
  \left[tr(\d\theta\d\theta) (\d\theta\d\theta)^{\mu\nu}\right]_{S^2\times S^2}
  &=  \frac 14 g^{\mu\nu} \left[tr(\d\theta\d\theta) tr(\d\theta\d\theta)\right]_{S^2\times S^2} \ .
\end{align}
Contracting with $\tilde T[\phi]_{\mu\nu}$ and
recalling  that $S_4[\d\theta] = 0$  for the self-dual background,
we conclude that 
\begin{align}
 S_4[\phi] &= 0\  + O(\phi^4) + O((\del^2\phi)^2) \ .
\end{align}
Therefore transversal deformations of the background 
do not acquire quadratic quantum corrections at one loop,
up to possible subleading higher-derivative terms.
As a check, $S_4[\phi]$  vanishes for radial deformations
$A^a = X^a$, where $\tilde T^{\mu\nu} \sim g^{\mu\nu}$.
This is in contrast to tangential deformations, as we will see.

\subsection{Tangential fluctuations}

The one-loop effective action is given by
\begin{align}
\Gamma_{\!\textrm{1loop};4}[X]\! 
  &= \frac 14 \frac{(\dim\cH)^2}{(\rm Vol(\cM^6))^2}\,
  \int\limits_{\cM\times\cM} d{\bf x} d {\bf y} 
   \frac{3 S_4[ \d\theta({\bf x,y})+\d\cF({\bf x,y})] }{(|x-y|^2+2\Delta^2)^4} \ .
   \label{gamma-S4-3}
\end{align}
The propagators act like a short-range delta function
with normalization
\begin{align}
  \int\limits_{\cM^4} dx  \frac{1}{(|x-y|^2+2\Delta^2_x)^4} 
 &\approx  \int\limits_{\R^4} d^4 x  \frac{1}{(|x|^2+2\Delta^2)^4} 
   =  \frac{\pi^2} {6}\, \frac{1}{\D^4}  \ 
\end{align}
using the Riemannian measure, and 
 $\cM^4$ indicates  $S^4$ with radius $R$.
Therefore the dominant contribution  will come from local interactions with 
${\bf x} = (x,\xi)$ and ${\bf y} = (x,\xi)$  denoting the same $x\in \cM^4$ 
but different points on the internal $S^2$. We can therefore replace
the integral over $\cM^4\times \cM^4$ by a single integral as follows
\begin{align}
\Gamma_{\!\textrm{1loop};4}[X]\! 
  &= \frac {\pi^2}{8} \frac{(\dim\cH)^2}{(\rm Vol(\cM^4))^2(\rm Vol(S^2))^2}\, \frac{1}{\D^4} 
 \!\!\! \int\limits_{\cM^4\times S^2\times S^2} \!\! d x d\xi d\eta\,  
    S_4[\d\theta({\bf x},{\bf y})+\d\cF({\bf x},{\bf y})] \ .
   \label{gamma-S4-4}
\end{align}
For the tangential fluctuations, $S_4$ can be written locally using \eq{F+F-indentity}
in terms of (anti-selfdual) flux components as follows:
\begin{align}
 S_4[\d\theta+\d\cF] 
  &= - 4 \big(\theta_+({\bf x})-\theta_+({\bf y}) + \cF_+({\bf x})-\cF_+({\bf y})\big)^{\mu\nu}
  \big(\theta_+({\bf x})-\theta_+({\bf y})+\cF_+({\bf x})-\cF_+({\bf y})\big)_{\mu\nu} \cdot\, \nn\\
  &\qquad \cdot \big(\cF_-({\bf x})- \cF_-({\bf y})\big)^{\r\s}{\big(\cF_-({\bf x})-\cF_-({\bf y})\big)}_{\r\s} \nn\\
  &\approx - 4 m^2({\bf x},{\bf y})\big(\cF_-({\bf x})-\cF_-({\bf y})\big)^{\r\s}
  {\big(\cF_-({\bf x})-\cF_-({\bf y})\big)}_{\r\s} \leq 0 , \nn\\
 \cF_\pm &= \cF \pm \star_g \cF  \nn\\
 m^2({\bf x},{\bf y}) &= (\theta({\bf x})-\theta({\bf y}) )^{\mu\nu}(\theta({\bf x})-\theta({\bf y}))_{\mu\nu} 
 \ \sim \  \D^4  \|\xi-\eta\|^2  \ > 0
 \label{S_4-tang-general}
\end{align}
using \eq{m2-S2}.
Here we used the self-duality of the background  $\theta_-=0$, while 
$\theta_++\cF_+ \approx \theta_+$ for small fluctuations.
As above, ${\bf x,y}$ denote the same $x\in \cM^4$ 
but different points $\xi,\eta$ on the internal  $S^2$.
This should be integrated  over $S^2\times S^2$ for each $x\in\cM^4$. 
Since $m^2 >0$ whenever $\xi\neq \eta$, only the ASD components $\cF^{\mu\nu}_-$
 contribute, {\em with a negative sign}.
Hence fluctuations  $\cF(x) \in (n,0)$ which are constant along $S^2$
drop out, but all the higher spin fluctuations such as  
$\cF(x) \in (n,2)$ will contribute.

The gravity modes of interest here give rise to  $\cF^{\mu\nu} \in (n,2)$. These can be written as 
\begin{align}
 \cF^{\mu\nu}(x,\xi) = \cF^{\mu\nu}_a(x)  \xi^a 
\end{align}
for $\xi_a\in S^2$ and $\cF^{\mu\nu}_a$ a  3-vector.
Then 
\begin{align}
\big(\cF(\xi)-\cF(\eta)\big)^{\mu\nu}{\big(\cF(\xi)-\cF(\eta)\big)}_{\mu\nu} 
  &=  \cF^{\mu\nu}_a  \cF_{\mu\nu}^b (\xi-\eta)_a (\xi-\eta)^b  \nn\\
  \cF_a^{\mu\nu} \cF^a_{\mu\nu} &= 3\big[\cF^{\mu\nu}(\xi) \cF_{\mu\nu}(\xi)\big]_{S^2} \ .
\end{align}
Using 
\begin{align}
 \frac 1{({\rm Vol} S^2)^2} \int\limits_{S^2\times S^2} \|\xi-\eta\|^2   (\xi-\eta)^a (\xi-\eta)^b 
= \frac{16}{9} \d^{ab}
\end{align}
 we can write
\begin{align}
 \frac 1{({\rm Vol} S^2)^2} \int\limits_{S^2\times S^2}  \|\xi-\eta\|^2
 \big(\cF(\xi)-\cF(\eta)\big)^{\mu\nu}{\big(\cF(\xi)-\cF(\eta)\big)}_{\mu\nu} 
   =  \frac{16}{3} \big[\cF^{\mu\nu}(\xi) \cF_{\mu\nu}(\xi)\big]_{S^2} \ . 
\end{align}
This means that for such $\cF\in (n,2)$, the 1-loop effective action at $O(\cF^2)$ can be written as 
\begin{align}
 \Gamma_{1-loop;4}[\cF^2]  
  &= -\frac{8\pi^2}3\frac{(\dim\cH)^2}{(\rm Vol\cM^4)^2}\,
  \int\limits_{\cM^4} dx\, \big[\cF^{\mu\nu}_-(\xi) \cF_{\mu\nu}^-(\xi)\big]_{S^2} 
\end{align}
where 
\begin{align}
 \big[\cF_-^{\mu\nu} \cF^-_{\mu\nu}\big]_{S^2} 
 &= \big[2\cF^{\mu\nu} \cF_{\mu\nu}  -\cF^{\mu\nu} \cF^{\r\s} \epsilon_{\mu\nu\r\s}\big]_{S^2} \ .
\end{align}
As shown in section \ref{sec:eom}, 
this can be absorbed in a renormalized action \eq{fluct-action-bare-explicit} for a suitable value of $\a\neq 1$.

Clearly the maximal supersymmetry of the model protects the flat limit $R\to\infty$ from 
large quantum corrections (i.e. from the non-local UV/R mixing), leading only to the above mild term. 
Note that there is no ``cosmological constant'' induced at one loop; in fact the very concept does not apply 
in this framework, which is based on  matrix degrees of freedom rather than a fundamental metric. 
Only the  background curvature (which we dropped) might lead to modifications in the linearized  
Einstein equation \eq{Einstein} which look like a cosmological constant.
Hence the ``cosmological constant'' problem is replaced here by the question of stability of a background 
with sufficiently large $R$ and small extra dimensions. These are hopefully 
feasible problems which need to be addressed in future work.

\section{Conclusion and outlook}

We have shown that the 4-dimensional 
(Euclidean, linearized) Einstein equations  emerge from 
the  dynamics of fluctuations on  fuzzy 4-spheres $\cS^4_\L$  in  
Yang-Mills matrix models, in a certain regime and provided certain conditions are met.
The resulting  physics is  richer than in general relativity, since 
there are several contributions to the metric.
Most importantly, gravity is cut off at some long-distance scale $m^{-1}$ or $L_R/\a$.
Moreover, a tower of higher-spin fields  arises on top of the gravitational 
modes, leading to a higher-spin theory. 
The present analysis is expected to capture the leading gravitational effects, 
since fields with spin larger than 2 should decouple at low energies.
Thus the gravitational physics of the present model could be  
sufficiently close to general relativity at least for solar-system scales.

The conditions to obtain an interesting gravity are as follows: 1) 
the background must be a generalized "thick"  fuzzy sphere $\cS^4_\L$ with $n \gg\sqrt{N}$,  
leading to a large scale $L_R/\a$ 
which acts as an IR cutoff for gravity, and 
2)  dimensional reduction to 4 dimensions is justified. 
We discussed possible mechanisms for the latter.
One obvious mechanism involves the radial potential which stabilizes $\cS^4_N$. Another
possibility is to give VEV's to the transversal scalar fields 
 along the lines of \cite{Steinacker:2014lma,Steinacker:2014fja}, leading to fuzzy extra dimensions.
 This is  natural given the structure of $\cS^4_\L$ as bundle over $\cS^4_N$, and
 it would also provide an interesting symmetry breaking structure,
 leading to a low-energy gauge theory in the right ball-park of particle physics
 \cite{Steinacker:2015mia}.  
 Yet another  possibility is to have a self-dual Yang-Mills action\footnote{There are indeed hints that this arises
 taking fully into account the volume fluctuations, cf. \cite{Steinacker:2010rh,Steinacker:2007dq}.} 
 \eq{action-SD}; then $\a=0$, and the extra dimensions (i.e. $c_n$) may be  small
 (but non-zero; the basic fuzzy sphere $S^4_N$ does not suffice). 
 Anyway, it is intriguing that the generalization to $\cS^4_\L$ seems to provides
 the required ingredients for both gravity and interesting particle physics.

 To clarify these conditions requires a more detailed treatment of the generic fuzzy spheres $\cS^4_\L$
(cf. appendix \ref{sec:app-geom-general}), as well as an understanding of the effective potential
for the extra dimensions which would arise at one loop.
Assuming that these conditions can be met,  
the long-wavelength modification of gravity discussed above could be very interesting, as
they might lead to behavior usually attributed to dark matter or dark energy. 
There will also be  new effects due to additional modes arising e.g. from radial deformations $\k$.

Apart from the above conditions, 
there are other issues which need to be addressed before  physical implications can be extracted. 
One is to find a suitable Minkowski version of the background.
While most of the analysis will generalize,  the proper 
choice of a covariant Minkowskian matrix geometry is not clear, and
there are  non-trivial issues related to the non-compactness of 
the Lorentz group\footnote{One problem is that the internal fiber could be noncompact
as already noted in \cite{Doplicher:1994tu}, hence the meaning of averaging 
is not clear. However, the expansion into higher spin modes would still go through.}.
Natural candidates would be based on a non-compact version of 
$SO(6)$ (cf. \cite{Heckman:2014xha}), or possibly some fuzzy de Sitter space \cite{Gazeau:2009mi,Buric:2015wta}.

The restriction to linearized gravity in this paper is clearly not essential.
The model is fully non-linear, and 
much of the derivation would go through for perturbations on a non-trivial background.
We simply have to make the replacement \eq{cA-expand-e} in the general mode expansion \eq{cA-expand-4D},
and perturbations around a non-trivial $\g^{\mu\nu}$  
could be studied along the same lines, leading presumably to the full  Einstein equations on $\cS^4_\L$. 
Hence there is no obstacle for describing strong gravity in this manner.

 For the IKKT matrix model, the quantization should be well-behaved,
 and the present mechanism provides a promising basis for a quantum theory of gravity 
 with low-energy physics close to GR. The maximal supersymmetry  protects  
 backgrounds with large radius, and leads to a stabilization \cite{Steinacker:2015dra}. Moreover the 
 non-local UV/IR mixing is mild in this model, and reduces to 10-dimensional supergravity 
 in the bulk  \cite{Ishibashi:1996xs,Douglas:1998tk,Steinacker:2016xx}.
 We have started this endeavour by computing the leading one-loop corrections for the simplest  
 fuzzy 4-sphere, which lead to modified parameters of the action including $\a$.

 The relation of the IKKT model with IIB string theory also suggests an interesting general message: 
 compactification of target space may not be needed,
 so that the vast landscape of string compactifications may be avoided. 
 While IIB supergravity arises in the bulk upon quantization, 
 this has {\em nothing} to do with the present mechanism for gravity, which is purely classical. 
 The present mechanism should therefore not be confused with mechanisms to 
 localize bulk gravity to the brane such as \cite{Randall:1999vf}.
 If it is possible to obtain also a (near-) realistic low-energy particle physics in this framework
 (e.g. along the lines of \cite{Steinacker:2015mia,Steinacker:2014lma,Steinacker:2014fja}),
 it would provide an extremely simple and attractive approach to a quantum theory of fundamental interactions.

\paragraph{Acknowledgements.}

I would like to thank  J. Barrett, S. Fredenhagen, M. Hanada, J. Karczmarek, H. Kawai for useful discussions, 
and S. Ramgoolam and J. Zahn for related discussions and collaboration.
This work was supported by the Austrian Science Fund (FWF) grant
 P28590, and by the Action MP1405 QSPACE from the European Cooperation in Science and Technology (COST).

\appendix

\section{The classical geometry of the 4-spheres $\cS^4_\L$}
\label{sec:app-geom-general} 

The fuzzy 4-spheres under consideration are quantizations of the (co)adjoint orbits 
$\cO[\L] = \{g\cdot H_\L \cdot g^{-1}; \ g\in SU(4)\} \hookrightarrow \msu(4)$ 
projected to $\R^5$ via the projection $\Pi$ \eq{X-embed-Pi},
\begin{align}
 \cS^4_\L := \Pi(\cO[\L]) \ \subset \ \R^5 \ .
\end{align}
The coadjoint orbit is a homogeneous space $\cO[\L] \cong SU(4)/\cK$ where $\cK$ is the stabilizer of $\L$.
Here we discuss the classical geometry of these spaces and their harmonics.
This is best understood in terms of 
the spinorial representation of $\msu(4) \cong \mso(6)$ on $\C^4$.
Let  $\gamma_{a}$ be $4\times 4$ hermitian gamma matrices of $SO(5)$
with $\{ \gamma_{a},\gamma_{b}\}=2 g_{ab}$ for $a, b = 1,...,5$.
To be specific, we  choose the Weyl basis where 
\begin{align}
 \g_5 = \begin{pmatrix}
              \one_2 & 0\\
              0 & -\one_2
             \end{pmatrix} \ .
\label{gamma-5}
\end{align}
Then a 4-dimensional representation of $\mso(6)$ can be defined by the following 
generators \cite{Govil:2013uta}
\begin{align}
\Sigma_{\mu\nu} := \frac{1}{4i}\left[\gamma_{\mu},\gamma_{\nu}\right]
\qquad
\Sigma_{\mu 5} :=  -\frac{i}{2}\gamma_{\mu}\gamma_{5}
\qquad
\Sigma_{\mu 6} :=  -\frac{1}{2}\gamma_{\mu}
\qquad
\Sigma_{56} := - \frac{1}{2}\gamma_{5} \ 
\label{sigma-explicit}
\end{align} 
where $\mu,\nu=1,...,4$.
The embedding of $\cO[\L] \hookrightarrow \mso(6) = \R^{15}$ is then described by 
the 15 (real-valued, commutative) embedding functions 
\begin{align}
 m^{ab} &= tr(\Xi\, \Sigma^{ab}) , \qquad a,b=1,...,6 , \qquad \Xi \in \ \cO[\L].
  \label{mab-def-class}
\end{align}
The point $\Xi \in \cO[\L]$ can then be recovered from 
\begin{align}
 \Xi = \sum_{1\leq a<b\leq 6} m^{ab} \Sigma_{ab} \qquad  \in \ \cO[\L] \ .
 \label{zeta-reconstruct}
\end{align}
In particular,
\begin{align}
 x_a &= tr(\Xi \Sigma_{a6})  = -\frac 12 tr(\Xi\g_a)  
\end{align}
(setting $r=1$)
defines the embedding of $\cO[\L]$ in $\R^5$, which is the classical limit of  $\cS^4_\L$.

The corresponding quantized (``fuzzy'') coadjoint orbits are simply obtained by replacing the functions $m^{ab}$
on $\cO[\L]$ by the generators $\cM^{ab}$ acting on the highest weight irrep  $\cH_\L$, where
$\L$ should be a (dominant) integral weight. More details can be found e.g. in
\cite{Hawkins:1997gj}.

\subsection{The basic  sphere $\cS^4_N$}

The fuzzy sphere $\cS^4_N$ is obtained for $\L = N \L_1 = (0,0,N)$ or equivalently\footnote{For better 
readability we  do not impose
tracelessness here. This does not lead to significant changes.}
\begin{align}
 H_N \equiv H_{N\L_1} = N|\psi_0\rangle\langle \psi_0| \qquad \mbox{for} \ \ |\psi_0\rangle = (1,0,0,0)^T\in \C^4
\end{align}
The stabilizer of $\L$ is
$\cK=SU(3)\times U(1)$, and clearly $\cO[\L] \cong \C P^3$. 
By inspection of \eq{gamma-5} we find 
\begin{align}
 x_\mu &=  -\frac 12 tr(H_\L\g_\mu)  = 0, \qquad \mu=1,...,4 \nn\\
 x_5 &= \frac N2 = R_N \ .
\end{align}
This defines our  reference point  $x_{(0)} \in S^4$ (the ``north pole'').
It is easy to see using $SO(5)$ invariance and the explicit form of the generators \eq{sigma-explicit} that 
\begin{align}
  x_a x^a \ &\equiv \sum_{a=1}^5\ x_a^2\  = R_N^2 \nn\\
  m_{ab} &= \frac 1{2R}\, \epsilon_{abcde} m^{cd}  x^e \nn\\[1ex]
 p_\mu &\propto m_{\mu 5} = tr(H_\L \Sigma_{\mu 5}) = 0 \ .
\end{align}
The second identity expresses self-duality.
The stabilizer group of $x_{(0)}$ is
\begin{align}
 \{h \in SO(5); [h,\g_5] = 0\} \ = SU(2)_R\times SU(2)_L   \subset SO(5) 
\end{align}
where $SU(2)_L$ acts on the $+1$ eigenspace of  $\g_5$. 
Hence there is a fiber of points ${\bf x}\in\C P^3$ over each point $x\in S^4$,
which at the reference point $x_{(0)}$ is obtained by acting with 
$SU(2)_L$ on $|\psi_0\rangle$. These fibers are resolved by 
the functions $m_{\mu\nu},\ \mu,\nu=1,...,4$, which
define a tangential SD rank 2 tensor (or a 2-form) on $S^4$ with
\begin{align}
  m_{\mu\nu}  m^{\mu\nu} &= 4R_N^2 \ .
\end{align}
These define  2 independent functions, which 
describe the internal $S^2$ fiber of $\cS^4_N \cong \C P^3$ over $S^4$.
However, the  ``momentum'' functions $p_\mu$
vanish for any point on the fiber over $x$. 
Hence there are no independent modes $F_\mu(x) p^\mu$ 
on the basic sphere $\cS^4_N$. 
Another way to see this is via the Poisson bracket identity 
\begin{align}
 0 = \{x^b x_b,x^a\} = 2x_b m^{ba}
\end{align}
since $x_b x^b = R_N^2$ for the basic fuzzy 4-sphere (but not for the generalized ones).
At the north pole, this gives $p^\mu = 0$.
Moreover, the following identity of $\mso(6)$ tensors holds 
\begin{align}
 \sum_{a=1}^6 m^{ab} m^{ac}
 \equiv \sum_{\mu=1}^4 m^{\mu a} m^{\mu b} +  x^a x^b = R_N^2 \d^{ab} \ . 
 \label{mm-id-basicS4}
\end{align}
Besides direct verification, this follows (similar as in section \ref{sec:S4-properties})  from the fact   that
$\cC^\infty(\C P^3)$ does not contain any $(0,2,0)$ modes, leaving only the trivial tensor $\d^{bc}$ for the rhs.

\subsection{The generalized  sphere $\cS^4_\L$}

Now consider 
$\cS^4_\L$ for $\L = N{\L_1} + n_1\L'_{1} + n_2\L'_{2} $, 
where $\L'_i$ are fundamental weights of the $\msu(3)$ stabilizator of $\L_1$ (hence orthogonal to $\L_1)$,  
for $n_1, n_2 \ll N$. Then
\begin{align}
 H_\L =  H_{N\L_1} + H'_{n_1,n_2} &= N |\psi_0\rangle\langle \psi_0| 
 + \sum_{i=1,2}  n_i |\psi_i\rangle\langle \psi_i|,  \nn\\
 \langle \psi_i|\psi_j\rangle &= \d_{ij} \ .
\end{align}
Let $P$ be the spectral function of $H_\L$ which maps the small eigenvalues $n_i$ to zero.
Then 
$P$ (extended to the entire $SU(4)$ orbit)  
 projects  $\cO[\L]$ to $\cO[N\L_1] \cong \C P^3$.
Geometrically, this means that 
the generic orbits $\cO[\L]$ are naturally  
bundles over $\C P^3$, 
\begin{align}
 &\cO[\L]&  \nn\\ 
 &\ P \downarrow& \nn\\ 
 &\cO[N\L_1] \cong \C P^3 \ \stackrel{x^a}{\to}\  S^4 \ \hookrightarrow \ \R^5 \ .
\end{align} 
The fibers of this bundle are given by the $\msu(3)$ coadjoint orbits 
$\cO_n :=\{U H'_{n_i} U^{-1}, \ U \in SU(3)\}$, which
are resolved by the functions $p^\mu$ and $m^{\mu\nu}$ on $\cO[\L]$. 
More precisely, for $n_2=0$ this is the 4-dimensional space $\cO_n \cong\C P^2$ parametrized by $p^\mu$, while
$m^{\mu\nu}$ is still self-dual and describes the $S^2$ fiber of $\C P^3$ over $S^4$.
For  $n_1,n_2\neq 0$, $\cO_n$ is a 6-dimensional coadjoint orbit of $\msu(3)$ parametrized by $p^\mu$
and the ASD components of $m^{\mu\nu}$. For simplicity we assume $n_2=0$, and 
$H_\L = H_{N\L_1} + H'_n$ where $H'_n = U diag(0,n,0,0)U^{-1}$ for $U\in SU(3)$.
Then 
\begin{align}
 m^{\mu 5} &= tr(H_\L \Sigma_{\mu 5}) = tr(H'_n  \Sigma_{\mu 5})
 =  -\frac i2 tr\begin{pmatrix}
                  0 \\
                    & U\begin{pmatrix}
                      n & \\ & 0 & \\ && 0
                       \end{pmatrix}U^{-1}
                  \end{pmatrix}
   \begin{pmatrix}
    0 & \sigma^\mu\\
    \tilde\sigma^\mu & 0
   \end{pmatrix} \neq 0
\end{align}
which is not constant along $\cO_n$. Upon averaging over the local fiber, one obtains
\begin{align} 
 [m_{\mu 5} m^{\mu 5}]_{0} = c_n^2  = O(n^2) > 0
 \label{PP-normalization}
\end{align} 
(which we refrain from computing here explicitly).
Hence in contrast to  the basic $\cS^4_N$, the  ``momentum'' functions $p^\mu$ are independent, 
so that  the modes  $F_\mu(x) p^\mu$ are non-trivial.
Similarly, the radial function 
\begin{align}
 \cR^2= x_a x^a &= \frac 14 tr(\Xi\otimes \Xi \ \g_{a} \otimes \g^{a}) 
  =  \frac 14 tr\big(\Xi\otimes \Xi \ (- \one + 2P + 8 P_1)\big) \nn\\
  &= \frac 14 \big(-(tr \Xi)^2 +2 tr(\Xi^2) +16 N n (\epsilon \bar \psi_0 \bar\psi_1) (\epsilon\psi_0  \psi_1)  \big) \nn\\
  &=  \frac 14 \big(-(N+n)^2 +2(N^2 + n^2)  + 16N n |\tau|^2 \big) ,
          \qquad \tau =  \epsilon\psi_0  \psi_1 \ \in \C \nn\\
 &=  \frac 14 \big(N^2 - N n +n^2 + 16N n |\tau|^2 \big) ,
          \qquad \tau =  \epsilon\psi_0  \psi_1 \ \in \C 
\label{Rsquare-Z-class}
\end{align}
for $\Xi\in\cO[\L]$ using \eq{gamma-tens-id},
where $P$ is the permutation operator acting on $\C^4 \otimes \C^4$.
Now the point is that $\t$ is {\em not} invariant under $SO(6)$, so that the spectrum of
$\cR^2$ lies in an interval $[R^2_{min},R^2_{max}]$ peaked around $R_N^2 =\frac{N^2}{4}$.
This means that the generic 4-spheres $\cS^4_\L$
are ``thick'' spheres, with $\{\cR^2, x^a\} \neq 0$. This essential for 
the existence of independent momentum functions $p^\mu$, which are the basis of the present mechanism for gravity.
Finally we note that the identity \eq{mm-id-basicS4} still holds approximately, in the form
\begin{align}
\sum_{\mu=1}^4 m^{\mu a} m^{\mu b}  
 = R_N^2\, \Big(P_T^{ab} \ +   t^{ab}   \Big)
 \label{mm-id-generalS4}
\end{align}
where $P_T^{ab}= \d^{ab} - \frac 1{R_N^2}x^a x^b$ is the tangential projector on 
$S^4 \subset \R^5$, and $t^{ab} = O\big(\frac{n}{N}\big)$ arises from  $(0,2,0)$ 
modes in $\cC^\infty(\cS^4_\L)$.

\section{Some identities for fuzzy 4-spheres}
\label{sec:S4-properties}

First, we note the following identity for the $SO(5)$ gamma matrices 
\begin{align}
  \g_a \otimes \g^a 
  &= \frac 12 (\one+P) -\frac 32 (\one-P) + 8 P_1 \ .
  \label{gamma-tens-id} 
\end{align} 
Here $P_1 = \bar \epsilon \epsilon$ is the projector on the $\mso(5)$ singlet in 
$(4)\otimes (4) = ((10)_S \oplus (6)_{AS}\big)_{\mso(6)} 
  = ((10)_S \oplus (5)_{AS} \oplus (1)_{AS}\big)_{\mso(5)}$, 
which is broken by $\mso(6)$. 
Furthermore, we are interested in the following tensor operator 
\begin{align}
 \cT^{ab} := \frac 12\sum_{a,a'=1}^6 \{\cM^{ab},\cM^{a'c}\}_+ \d_{aa'} \ .
\end{align}
Consider first  
\paragraph{The basic fuzzy sphere $\cS^4_N$.} 
Since $End(\cH)$ does not contain any 
$(0,2,0)$ modes, it follows\footnote{Note that $(0,1,0)$ is the 6-dimensional vector representation of $\mso(6)$.} 
that $\cT^{ab} \sim \d^{ab}$.
Computing the trace $\cT = \cT^{ab}\d_{ab} =  2 C^2[\mso(6)] = \frac 32N(N +4)$ 
(cf. \cite{Steinacker:2015dra}), we obtain 
\begin{align}
 \cT^{ab} 
 = \frac 13 C^2[\mso(6)]\d^{bc} = R_N^2\d^{bc}
 \label{MM-6-basic}
\end{align}
i.e. 
\begin{align}
  \frac 12\sum_{a,a'=1}^5 \{\cM^{ab},\cM^{a'c}\} g_{aa'} &= R_N^2 g^{bc} -\frac 12 \{X^b, X^c\} \ .
\end{align}
This is the fuzzy analog of \eq{mm-id-basicS4}.
For the
\paragraph{Generalized fuzzy spheres} $\cS^4_\L$
with $\L = (n_1,n_2,N)$, 
 $End(\cH)$ may contain some $(0,2,0)$ modes. Then 
the above relation generalizes as 
\begin{align}
 \frac 12\sum_{a,a'=1}^6 \{\cM^{ab},\cM^{a'c}\} \d_{aa'} = \frac 13 \d^{bc} C^2[\mso(6)] + t^{ab}
  \label{MM-6-generic}
\end{align}
where $t^{ab}$ is a traceless $(0,2,0)$ tensor operator of order
$t^{ab} = O(n) \ll C^2[\mso(6)]$, which is suppressed. This is the fuzzy analog of \eq{mm-id-generalS4}.

\stoptocwriting

\section{Background flux $\theta^{\mu\nu}(x,\xi)$ averaged over $S^2$}

We need various averages of the 
background flux $\theta^{\mu\nu}(x,\xi)$ over $S^2$. 
One useful result which follows from the self-duality and \eq{MM-formula-metric} is
  \begin{align}
  \left[ \theta^{\mu\nu} \theta^{\r\s}\right]_{0} 
 &= \frac{\D^4}{12} (\d^{\mu\r}\d^{\nu\s} - \d^{\nu\r}\d^{\mu\s} + \varepsilon^{\mu\nu\r\s}) \ .
  \label{M-correlators-2sphere}
\end{align}
This also applies to $\cS^4_N$, and to $\cS^4_\L$ as long as $N \gg n_i$.
Furthermore since $\theta^{\mu\nu}$ is self-dual, we can write  
\begin{align}
 \theta^{\mu\nu}(\xi) &=  r^2\theta^{\mu\nu}_a J^a(\xi)
\end{align}
where $J^a$ are the generators of the internal fuzzy sphere $S^2_{N+1}$, 
which in the semi-classical limit are functions $J^a: S^2 \to \R^3$ 
on $S^2$ with radius given by
\begin{align}
 \theta^{\mu\nu} \theta_{\mu\nu} = 4 r^4 J^a J_a  \sim N^2 r^4
\end{align}
using
\begin{align}
 \theta^{\mu\nu}_a \theta_{\mu\nu}^b = 4 \d_a^b  .
\end{align}
Therefore 
\begin{align}
  m^2(\xi,\eta) &= (\theta^{\mu \nu}(\xi)-\theta^{\mu \nu}(\eta)) (\theta_{\mu\nu}(\xi)-\theta_{\mu\nu}(\eta)) \nn\\
  &= 4 r^4(J^a(\xi)-J^a(\eta))(J_a(\xi)-J_a(\eta))  \nn\\
  &\sim \D^4\|\xi-\eta\|^2 
  \label{m2-S2}
\end{align}
where $\xi,\eta$ are unit vectors on $S^2$, and recalling $N^2 r^4 = \D^4$.

\section{Mixed Young projections and permutations}
\label{sec:young-projectors}

Define
\begin{align}
   P_{\rm hor} &:= \frac{1-P_{23}}2 P_{12}\frac{ 1-P_{23}}2 \nn\\
  P_{\rm hor}^2 &= - \frac 12 P_{\rm hor} +(1-P_{23}) \ .
   \label{P-equation}
\end{align}
Acting on tensors which are anti-symmetric in the last two indices we have  $P_{23}=-1$, and
\begin{align}
 (P_{\rm hor}+1)(P_{\rm hor}- \frac 12) &= 0 \ .
\end{align}
Hence solutions of $P_{\rm hor} = -1$ are the totally anti-symmetric Young diagrams,
while the solutions of $P_{\rm hor} = \frac 12$ are mixed (hook) Young diagrams $A_{\mu\r\s}$.
This means that interchanging the first two (``horizontal'') indices of such  $A_{\mu\r\s}$ costs a factor $\frac 12$.

%

\section{Evaluation of $D^2$}
\label{sec:eval-D2}

First, one easily derives from the basic $\cS^4_N$ algebra 
the following semi-classical results
\begin{align}
 \Box \theta^{\mu\nu} &= 2 r^2 \theta^{\mu\nu} , \qquad \Box P^\mu = 2 r^2 P^\mu \nn\\
 \Box (\theta^{\mu\nu} P^\s) 
  &= \Box \theta^{\mu\nu} P^\s +\theta^{\mu\nu}  \Box P^\s
  + 2 [X^\a,\theta^{\mu\nu}] [X_\a,P^\s] = 4 r^2\theta^{\mu\nu} P^\s   \nn\\  
  \Box (\theta^{\mu\nu} \cM^{\s\r}) &= 4 r^2\theta^{\mu\nu} \cM^{\s\r}  \nn\\
 2i[\theta^{\mu\mu'}, \theta^{\mu'\nu} P^\s] 
 &= 4r^2\theta^{\mu\nu} P^\s +2 \cM^{\s\nu}P^{\mu}  \nn\\
 2i[\theta^{\mu\mu'}, \theta^{\mu'\nu} \cM^{\s\r}] 
  &= -2r^2\big(g^{\mu\mu'}\theta^{\mu'\nu} -  g^{\mu'\mu'}\theta^{\mu\nu}+ g^{\mu'\nu}\theta^{\mu\mu'})\cM^{\s\r} \nn\\
  &\qquad -2 \theta^{\mu'\nu}\big(g^{\mu\s}\theta^{\mu'\r} - g^{\mu\r}\theta^{\mu'\s} 
             - g^{\mu'\s}\theta^{\mu\r} + g^{\mu'\r}\theta^{\mu\s}\big)  \nn\\
  &= 4r^2\theta^{\mu\nu} \cM^{\s\r} -2 \big(\g^{\nu\r} g^{\mu\s} - \g^{\nu\s} g^{\mu\r}
   - \theta^{\mu\r}\theta^{\s\nu} +\theta^{\mu\s} \theta^{\r\nu}\big)\nn\\
  &= 2(2\theta^{\mu\nu} \theta^{\s\r}  +\theta^{\mu\r}\theta^{\s\nu} -\theta^{\mu\s} \theta^{\r\nu})
  +2 \big(g^{\mu\r} \g^{\nu\s} -g^{\mu\s}\g^{\nu\r}\big)
\end{align}
noting that $\theta^{\mu\nu}P_{\mu}= 0$ at $p$. 
As a check, consider\footnote{Note that $\Box (\theta^{\mu\nu} \cM^{\s\r})$ is not consistent with a contraction by 
$g_{\nu\s}$, i.e. its trivial component $[\theta^{\mu\nu} \cM^{\s\r}]_{S^2}$ would require to
keep sub-leading terms.
However this correction is not significant.}
\begin{align}
  2i[\theta^{\mu\mu'}, \theta^{\mu'\nu}g_{\nu\s} \cM^{\s\r}] 
  &=  2(2\theta^{\mu\nu} \theta^{\s\r} -\theta^{\mu\s} \theta^{\r\nu})
  +2 \big(g^{\mu\r} \g^{\nu\s} -g^{\mu\s}\g^{\nu\r}\big)g_{\nu\s} \nn\\
  &=  2(-2\g^{\mu\r} -\g^{\mu\r}) +6 \g^{\mu\r} = 0 \ .
\end{align}
Using these results and the semi-classical rules \eq{semiclass-rules} 
we obtain
\begin{align}
 \Box (\theta^{\mu\nu} A_{\nu\s}(x) P^\s) &= [X^a,[X_a, \theta^{\mu\nu} A_{\nu\s} P^\s]] \nn\\
  &\sim (\Box+4r^2) A_{\nu\s} \theta^{\mu\nu} P^\s 
   +2\theta^{\mu\nu} \theta^{\a\s}\del_\a  A_{\nu\s} \nn\\
 \Box (\theta^{\mu\nu} A_{\nu\s\r}(x)\cM^{\s\r}) &=
 \Box A_{\nu\s\r} \theta^{\mu\nu} \cM^{\s\r} +  A_{\nu\s\r}\Box(\theta^{\mu\nu} \cM^{\s\r})
   +  2[X^a,\theta^{\mu\nu} \cM^{\s\r}][X_a, A_{\nu\s\r}]\nn\\
   &\sim (\Box +4r^2) A_{\nu\s\r} \theta^{\mu\nu} \cM^{\s\r}
\end{align}
always dropping terms like $[X^\a,\theta^{\mu\nu}] \sim x = 0$ at $p$, so that 
e.g. $[X^\a,\theta^{\mu\nu} P^\s] \sim -i\theta^{\mu\nu}g^{\a\s}$.

\resumetocwriting

\bibliography{papers}
\bibliographystyle{diss}

\end{document}